\documentclass[sigconf]{acmart}

\copyrightyear{2023}
\acmYear{2023}
\setcopyright{rightsretained}
\acmConference[CCS '23]{Proceedings of the 2023 ACM SIGSAC Conference on Computer and Communications Security}{November 26--30, 2023}{Copenhagen, Denmark}
\acmBooktitle{Proceedings of the 2023 ACM SIGSAC Conference on Computer and Communications Security (CCS '23), November 26--30, 2023, Copenhagen, Denmark}
\acmDOI{10.1145/3576915.3623174}
\acmISBN{979-8-4007-0050-7/23/11}

\makeatletter
\gdef\@copyrightpermission{
  \begin{minipage}{0.3\columnwidth}
   \href{https://creativecommons.org/licenses/by-nc-nd/4.0/}{\includegraphics[width=0.90\textwidth]{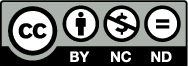}}
  \end{minipage}\hfill
  \begin{minipage}{0.7\columnwidth}
   \href{https://creativecommons.org/licenses/by-nc-nd/4.0/}{This work is licensed under a Creative Commons Attribution-NonCommercial-NoDerivs International 4.0 License.}
  \end{minipage}
  \vspace{5pt}
}
\makeatother

\usepackage{tikz}
\usepackage{amsmath}
\usepackage{graphicx}
\usepackage{textcomp}
\usepackage{xcolor}
\usepackage{caption}
\usepackage{subcaption}
\usepackage{amsmath}
\usepackage{algorithm}
\usepackage[noend]{algpseudocode}
\usepackage{glossaries}
\usepackage{tabularx}
\usepackage{mfirstuc}
\usepackage{multirow}
\usepackage{pgfplots}
\usepackage{pgfplotstable}
\usepackage{hyperref}
\usepackage{color}
\usepackage{balance}
\usepackage{multicol}
\usepackage{nopageno}
\usepackage{ifthen}
\usepackage{balance}

\newacronym{PDG}{PDG}{program dependency graph}
\newacronym{API}{API}{application programming interface}
\newacronym{AUC}{AUC}{area under the curve}
\newacronym{JRE}{JRE}{Java Runtime Environmemnt}
\newacronym{CFG}{CFG}{control flow graph}
\newacronym{SVM}{SVM}{support vector machine}
\newacronym{LSTM}{LSTM}{long short-term memory}
\newacronym{TLS}{TLS}{Transport Layer Security}
\newacronym{IEC}{IEC}{increase in ease and convenience}
\newacronym{ROC}{ROC}{receiver operating characteristic curve}
\newacronym{PCA}{PCA}{principal component analysis}
\newacronym{IV}{IV}{initialization vector}
\newacronym{ReLu}{ReLu}{Rectified Linear Unit}
\newacronym{UI}{UI}{user interface}
\newacronym{JCA}{JCA}{Java Cryptography Architecture}
\newacronym{GCS}{GCS}{Google Custom Search}
\newacronym{CSE}{CSE}{Custom Search Engine}
\newacronym{ESR}{ESR}{Enriched Search Results}
\newacronym{YOE}{YOE}{Years of Experience}

\newcommand{\diff}[1]{\color{black}{#1}\color{black}}

\newboolean{isArxiv}
\setboolean{isArxiv}{true}

\graphicspath{ {./images/} }

\begin{document}

\title{The Effectiveness of Security Interventions on GitHub}



\author{Felix Fischer}
\affiliation{%
  \institution{Technical University of Munich}
  \city{Munich}
  \country{Germany}}
\email{flx.fischer@tum.de}
\orcid{0009-0003-0995-8813}

\author{Jonas Höbenreich}
\affiliation{%
  \institution{Technical University of Munich}
  \city{Munich}
  \country{Germany}}
\email{jonas.hoebenreich@tum.de}
\orcid{0000-0002-3797-5691}

\author{Jens Grossklags}
\affiliation{%
  \institution{Technical University of Munich}
  \city{Munich}
  \country{Germany}}
\email{jens.grossklags@in.tum.de}
\orcid{0000-0003-1093-1282}


\begin{abstract}
  In 2017, GitHub was the first online open source platform to show security alerts to its users. It has since introduced further security interventions to help developers improve the security of their open source software. In this study, we investigate and compare the effects of these interventions. This offers a valuable empirical perspective on security interventions in the context of software development, enriching the predominantly qualitative and survey-based literature landscape with substantial data-driven insights. We conduct a time series analysis on security-altering commits covering the entire history of a large-scale sample of over 50,000 GitHub repositories to infer the causal effects of the security alert, security update, and code scanning interventions. Our analysis shows that while all of GitHub's security interventions have a significant positive effect on security, they differ greatly in their effect size. By comparing the design of each intervention, we identify the building blocks that worked well and those that did not. We also provide recommendations on how practitioners can improve the design of their interventions to enhance their effectiveness.
\end{abstract}




\maketitle

\section{Introduction}
Code security has emerged as a rapidly growing concern in the field of usable security research. A major contributing factor to the production of insecure code by developers is the reuse of vulnerable resources, such as insecure open source code and third-party libraries~\cite{acar2017comparing,Acar:dev:2016,fischer2017stack, fischer2021effect, Fischer19, chen2019reliable}. Perhaps surprisingly, even AI assistants can provide insecure recommendations since they are trained using insecure source code. The implications of insecure code are significant and can lead to severe security breaches, making it crucial to address this issue.

In response, numerous strategies aimed at assisting developers in producing secure and functional code have been studied. These include utilizing security alerts in development environments, code libraries, and compilers to a greater extent. Additionally, more user-focused warning designs that incorporate recommendations, quick fixes, or code examples within development environments or directly from open source platforms have been explored~\cite{acar2017comparing,Acar:dev:2016, fischer2017stack, Fischer19, gorski2018developers, fischer2021effect, fischer2022nudging, nguyen2017stitch}. However, despite these extensive research efforts, the effectiveness of these interventions has primarily been evaluated in laboratory settings or online research studies, and their practical impact remains uncertain.

GitHub was the first platform to make practical use of the diverse insights from prior research by deploying three security interventions in stages. On November 16, 2017, the platform introduced security alerts for vulnerable dependencies. Two years later, on May 23, 2019, GitHub announced an additional intervention that provides automated updates for vulnerable dependencies. GitHub's more recent security intervention, which scans code to detect a wide range of dangerous Common Weakness Enumeration (CWE) vulnerabilities in popular programming languages, was deployed on September 30, 2020.

As an open source platform, GitHub provides the research community with an opportunity to evaluate the impact of scientific research-based interventions on code security in practice. Sufficient time has elapsed since GitHub implemented its interventions to enable the collection of meaningful data for investigating the effectiveness of these interventions. This presents an opportunity to identify successful building blocks, as well as any flaws in the current design that require improvements.

However, a challenge arises in evaluating the effectiveness of these interventions, as they have already been deployed and GitHub users have been introduced to the new security innovations. This makes it difficult to perform randomized experiments, where one group receives the treatment and another group does not. Such experiments can only be conducted internally by GitHub. 

To provide the research community with a means of assessing interventions on GitHub, we demonstrate how to apply counterfactual prediction to measure the causal effects of interventions that have already been deployed. Counterfactual prediction involves predicting how the dependent variable, i.e., security of repositories, would have evolved if an intervention that actually happened had not occurred. The difference between the observed outcome and the counterfactual prediction represents the causal effect of an intervention.

We used counterfactual prediction to examine the causal effects of all three security interventions, which included security alerts and updates as well as code scanning. Our analysis showed that all three interventions had positive effects on code security, but the size of the effects varied significantly. While the deployment of security alerts and updates was generally successful, resulting in a relative increase of security updates by more than 149\%, the code scanning intervention was not as impactful, leading to an increase of only 2\% in the amount of patched vulnerabilities. We compared the interventions and offered recommendations on how to improve them based on what worked well and what did not.

We summarize our key contributions and findings as follows:
\begin{itemize}
    \item We conduct a time series analysis on security-altering commits covering the entire history of a large-scale sample of 41,097 GitHub repositories and  infer causal relationships involving security alerts and security updates within this dataset.
    \item The presence of security alerts related to vulnerable dependencies had the most significant positive impact, resulting in an approximately 149\% increase in the number of security updates manually executed by users.
    \item Automated security updates for vulnerable dependencies led to a further enhancement, resulting in an additional 32\% increase in the number of security updates.
    \item The majority of these security improvements implemented in response to the interventions remained effective and operational over time, underscoring the resilience of the repositories' functionality in the face of these interventions. This observation implies that the interventions, as introduced, did not compromise or disrupt the repositories' essential functionality.
    \item We additionally analyze the source code within a historical dataset of 491,525 repository snapshots, both before and after undergoing code scanning interventions, to assess the presence of two highly critical vulnerabilities in JavaScript and Python.
    \item The code scanning intervention had a moderate impact, leading to a reduction of 2\% of CWE-079 vulnerabilities in JavaScript source code. The observed effects on the reduction of CWE-020 vulnerabilities in Python source code were found to be insignificant.
\end{itemize}

We structure our work as follows: we first discuss related work in Section~\ref{ref:related} and introduce the studied interventions in Section~\ref{ref:interventions}. We present our methodology on sampling and on measuring the effects of the interventions in Sections~\ref{sec:sec_alerts_and_updates}--\ref{sec:counter_factual}, and present our results in Section~\ref{sec:results}. We discuss our work in Sections~\ref{sec:confounders}--\ref{sec:discussion},  and offer concluding remarks in Section~\ref{sec:conclusions}.

\section{Related Work}
\label{ref:related}
In the following, we discuss related work with a primary focus on the GitHub platform.

On June 29, 2021, GitHub announced the Copilot assistant\footnote{\url{https://copilot.github.com/}}, which, based on a large AI language model provides autocompleting functionality during code development. The feature is primarily available via a subscription model. Of particular relevance for our work is a recent research paper by Pearce et al. \cite{pearce2022asleep} demonstrating that in security-relevant code development scenarios, Copilot frequently produces suggestions with vulnerabilities on the Common Weakness Enumeration list. Research on training an AI language model to more reliably produce secure code is orthogonal to our research agenda, but such work could lead to other concrete security interventions on GitHub.

GitHub Actions -- released to the public in December 2019 -- allows for the automation of repetitive steps during software development on the platform. Kinsman et al.   \cite{kinsman2021how} investigated this tool empirically. Of relevance to our work is that only 13 out of 708 defined actions (in their set of studied repositories) focus on security. However, they also demonstrate that the automation offered by GitHub Actions is received favorably by early adopters.

Angermeier et al. \cite{angermeir2021enterprise} investigated the overall adoption of tools in enterprise-driven repositories on GitHub that use continuous integration frameworks. In their sample of 8,423 projects, 765 (6.83\%) showed evidence of security-related tool use; in particular, 169 were found to use Dependabot, and at least 68 used CodeQL. Within the investigated sample, the security-related tools were the most frequently used, which further emphasizes the relevance of our work.

Kavaler et al. \cite{kavaler2019tool} study tool choice by developers on GitHub in the context of JavaScript. However, they focus on quality assurance more broadly, rather than security. In particular, their sample does not focus on the security interventions studied in our paper. Their work, however, offers interesting insights regarding tool adoption decision-making and tool abandonment.

Tan et al. \cite{tan2021github} conducted an applied course project focused on security improvements on GitHub. Students (based on detailed instructions) provided suggestions to open source projects on GitHub to fix identified security issues. Interestingly, 93 of 214 suggestions were indeed merged by developers, which showcases a general willingness by GitHub users to adopt unsolicited security advice. 

In contrast, Panichella et al. \cite{panichella2021wont} provide interesting insights into why issues reported to repository owners are marked as ``wontfix''. Close to 50\% of wontfix issues are not addressed directly because they are marked as ``already
implemented or not needed.'' Other leading reasons are that the reported issues are not relevant or do not actually constitute a bug. However, in about 11\% of the cases the issues are acknowledged as bugs, but they are considered as too expensive or difficult to fix etc. In those cases, automatically provided recommendations could be particularly helpful. In their sample of wontfix issues, only 0.6\% have been specifically labeled as security-relevant issues.

Several reports investigate the demographics, perceptions and interests as well as concerns (e.g., regarding problems in the community such as harassment) of GitHub users, e.g., \cite{Geiger17,GitHub17,takhteyev2010investigating}. The GitHub Open Source Survey \cite{Geiger17,GitHub17}, for example, reveals that users believe open source software to be more secure than proprietary software (58\%) and that they consider security to be extremely important or very important (86\%). However, the survey did not address (potential or actual) tool usage on GitHub.

The GitHub platform is also regularly used to recruit subjects for research studies in security \cite{Acar17Github, gorski2018developers} or software engineering contexts \cite{Borges18,Gousios15,Goyal18,Jiang17,Pinto16,Ramos16,Rastogi16,Saito16,Wu14} as well as gender and diversity matters \cite{Vasilescu15,Vasilescu15b}.

Taken together, several studies investigate the usage and impact of tools on GitHub; but none explore security-focused interventions on GitHub. Given the importance and popularity of GitHub as a software development platform for open source as well as commercial software, and the fact that GitHub serves as a key innovator in security-oriented tooling our work addresses an important gap in practically relevant software engineering and security research.




\section{Security Interventions}
\label{ref:interventions}
Since 2017, GitHub has consecutively introduced three major security interventions, namely security alerts, security updates and code scanning. While the first two interventions scan and update vulnerable dependencies used by the source code of each repository, the third intervention scans the source code itself to find security weaknesses. Below, we provide an overview of the interventions and explain how they communicate their findings to the user. 

\paragraph{\textbf{Security alerts \& updates.}}
On November 16, 2017, GitHub introduced security alerts. The intervention scanned the repository for dependencies that contain Common Vulnerabilities and Exposures (CVEs) and suggested known fixes from the GitHub community. \diff{Security alerts were enabled by default } and were shown to the administrator of the repository in the dependency graph settings.

Two years later on May 23, 2019, GitHub acquired Dependabot, which also provides a feature for security updates. For each vulnerable dependency, Dependabot tries to find a secure version and raises a pull request that automatically updates the dependency to the minimum version including the patch. This intervention offered several notification types, which are enabled by default. An email is sent when a new vulnerability with a critical or high severity is found. UI alerts are shown in the repository's file and code views if Dependabot finds any vulnerable dependencies. Whenever users push to a repository via command line, a callback displays a warning if there are any vulnerabilities found. Web notifications are shown in the user's inbox when a new vulnerability with a critical or high severity is found. \diff{Security updates were not enabled by default and had to be activated by users. }

Dependabot opens a pull request for a repository in order to perform the security updates. The user can then decide whether they want to accept the request by clicking the merge button in the pull request view. This is the only action that the user is required to perform. The remaining parts of the intervention run completely automated.

\diff{Note, security alerts and security updates are two different interventions and are investigated separately in the remainder of this paper.}

\paragraph{\textbf{Code scanning.}}
On September 30, 2020, GitHub launched code scanning. The security intervention is powered by CodeQL, a static code analysis tool, which is able to detect vulnerabilities listed by the Common Weakness Enumeration (CWE) for a range of popular programming languages. \diff{In contrast to security alerts, code scanning is not enabled by default}. Users need to activate it in the security view on GitHub. On each push to the repository, CodeQL triggers a scan of the default branch and any protected branches. Pull requests targeted against the default branch are scanned automatically as well. Additionally, a repository is scanned on a fixed schedule once a week by default. The schedule can be adjusted by the user.


Code scanning alerts are shown in the repository's security view. They highlight vulnerabilities in the code that triggered the alert and provide annotations such as severity and information on how to fix the problem. CodeQL additionally provides data flow analysis that illustrates how data flows through the code. This is supposed to help users to identify areas of code that leak sensitive information. 



\section{Security alerts and updates}
\label{sec:sec_alerts_and_updates}
This section outlines the data, sampling method and metrics that were used for the causal effect analysis of the security alert and update interventions.

\subsection{Sampling}
We compiled a random sample from the complete population of 178,995,108 repositories on GitHub, collected until April 2022. The sample consists of 100,000 repositories that apply JavaScript or Python. We decided to focus on these programming languages due to their popularity (about 27\% of pull requests on GitHub; second quarter 2022). In addition, JavaScript and Python were the two most commonly used programming languages (according to percentage of pull and push requests) between 2013 and 2021.\footnote{\url{https://madnight.github.io/githut/}} 
Within the sample, we were able to identify 41,097 repositories that utilized at least one dependency with a recorded vulnerability in their version history. However, it is worth noting that the presence of such a vulnerability does not necessarily indicate that these repositories utilized a vulnerable version of the dependency. To determine whether the used version was indeed vulnerable, we cross-referenced the publicly available vulnerability list, which is also utilized by Dependabot, GitHub's own detection system. This list also provides additional details about the patched versions and the severity of the vulnerabilities.

\subsection{Metrics}
\label{sec:metrics}
In order to test and compare the effectiveness of the security alerts and updates we used three different metrics.

\paragraph{\textbf{Updated vulnerable dependencies}} Updated vulnerable dependencies will show how many users have replaced a vulnerable dependency with an updated version that fixed the vulnerability. This allows measuring differences in the updating behavior before and after the intervention. 

\paragraph{\textbf{Revoked updates}} This metric captures the amount of vulnerable dependencies that initially have been updated by the user, but were then replaced by the vulnerable version again. This allows analyzing whether a security update likely broke the repository, for instance due to compatibility issues. We think this metric is particularly important since it measures the usability of the intervention. According to related work, we would expect a large fraction of users to ignore security alerts once they experienced that updates break the functionality of their code (see, for example, observations by Johnson et al.~\cite{johnson13} and Danilova et al.~\cite{danilova2020}).

\paragraph{\textbf{Committed vulnerable dependencies}} Measuring the amount of committed vulnerable dependencies will allow evaluating whether the intervention had educational effects on users. It can indicate whether users became aware of certain vulnerable dependencies and whether they started using updated versions in future software development scenarios.



\subsection{Time series data}
\label{sec:time-series-data}
To apply the metrics outlined in Section~\ref{sec:metrics}, we generated time series data by downloading all repositories in our sample. For each repository, we scanned through the commits pushed to the \diff{active main } branch, and extracted those that altered the \textit{npm} configuration file. This file maintains a record of all the JavaScript dependencies utilized by the repository. We then used this information to construct three time series.

Firstly, the time series of updated vulnerable dependencies contains all commits from all repositories where a previously committed vulnerable version of a dependency was updated with a version that does not contain a known vulnerability.

Secondly, the time series of revoked updates contains all commits from all repositories where these updated dependencies were replaced once again with the previous vulnerable version.

Lastly, the time series of committed vulnerable dependencies contains all commits from all repositories that added a vulnerable version of a dependency to the repository.

For each commit in each time series, we record the identified dependency and its version, the timestamp, and additional metadata such as the repository URL, commit hash, message, and author.

\section{Code Scanning}
\label{sec:code_scanning}
In this section, we present the data, sampling method and metrics that were used for the causal effect analysis of the code scanning analysis.

\subsection{Sampling}
We drew a random sample of 2.5 million repositories from the population of 178,995,108 repositories hosted on GitHub until April 2022. The random sample only includes repositories that utilize JavaScript and Python, the two most popular programming languages on GitHub. These languages have several vulnerabilities that are listed among the top five most dangerous software weaknesses, such as CWE-020 and CWE-79, which we studied in the analysis of the code scanning intervention.\footnote{\url{https://cwe.mitre.org/top25/archive/2021/2021_cwe_top25.html}}


To evaluate the effectiveness of code scanning, we focused on fixing detected instances of CWE-79, the most dangerous JavaScript vulnerability according to CWE, and CWE-020, one of the most dangerous Python vulnerabilities. To meet our criteria, our sample had to include repositories that applied JavaScript and Python, had code scanning activated, and had at least one commit since the activation of the intervention. We also limited our sample to repositories that applied scanning schedules to ensure that CodeQL was being run on a frequent basis.

We implemented several exclusion criteria to ensure the quality of our analysis. Firstly, we excluded repositories that were forks. When analyzing a snapshot of a fork, the results would reflect the state of the parent repository, which might skew results by increasing the weight of duplicated code. Secondly, repositories that contained copies of OWASP Juice Shop\footnote{\url{https://owasp.org/www-project-juice-shop/}} and WebGoat\footnote{\url{https://owasp.org/www-project-webgoat/}} were also excluded since these repositories serve educational purposes, some even with intentional security vulnerabilities. Thirdly, we excluded a strong outlier from the sample. The project fixed 1,200 vulnerabilities more than five years before CodeQL was enabled and did not show any further change that could have been attributed to CodeQL. The outlier strongly influenced the average and, therefore, distorted the results. By applying these exclusion criteria, we ensured that the remaining sample was more representative of actively maintained repositories that use CodeQL on a frequent basis. Finally, we made sure that our samples did not deactivate CodeQL at any point during the post-period.

To select the samples, we used the GitHub API to search for repositories that met certain criteria, such as having previously activated CodeQL and being actively maintained after the introduction of the code scanning intervention. Using a search parameter that allowed us to query for commits made during a specific day, we searched through all days since the deployment of the code scanning intervention on GitHub. When we found a commit that added the CodeQL configuration file to the repository, we added the repository to the sample. Out of the 2.5 million JavaScript/Python repositories in total, 37,856 met the criteria. From this sample, we randomly selected 10,000 repositories for analysis.

We can estimate that this sampling method allowed us to collect approximately 70\% of all JavaScript/Python repositories that used CodeQL. To estimate this amount, we randomly sampled a further set of 40,000 repositories that use JavaScript and Python from the complete GitHub population. 

Using Cochran's formula, we determined that the optimal sample size for estimating proportions of dichotomous variables is 37,823. Considering a population of approximately 180 million repositories, a 99.99\% confidence interval with a 1\% margin of error, and an estimated 50\% activation rate of CodeQL (note that this percentage maximizes the required sample size), our sample size of 40,000 is deemed sufficient.

We observed that only 10 (0.00025\%) of these repositories used CodeQL. Based on this result, we can extrapolate that from the complete population of 178,995,108 repos, at most 54,348 repositories would meet the same criteria. Therefore, our collected sample can be considered 70\% of the estimated complete set of repositories that meet the criteria.

\subsection{Metrics}
To accurately measure the effectiveness of the code scanning intervention, we analyzed multiple snapshots of the repositories in our sample and measured the number of detected vulnerabilities belonging to the CWE-079 and CWE-020 vulnerability classes. By comparing the number of detected vulnerabilities before and after the introduction of the intervention, we were able to determine the extent to which users addressed these vulnerabilities in the source code of their repositories. This approach enabled us to quantify the impact of the code scanning intervention on improving code security.

\subsection{Time series data} 
We utilized the following sampling interval to create a time series for each repository. Starting 500 days before the repository activated code scanning, snapshots were taken every 7 days, collecting data until 200 days after activation of code scanning. To reduce the amount of source code that needs to be analyzed, we have limited the length of the time series to 700 days. We are confident that a pre-period of 500 days and a post-period of 200 days are sufficient to identify any causal effects that may exist.

As we created a separate time series for each repository, we needed to merge and align them into a single time series that could be analyzed. We aligned the time series using the activation date of code scanning as the reference point, with timestamps given a positive or negative time distance to the activation date in days. However, note that merging time series data resulted in missing snapshots for some repositories, for which we were unable to report vulnerability information for particular timestamps during analysis. To address this, we applied interpolation on each time series to obtain results for the missing snapshots.

\subsection{CodeQL}
CodeQL is an open source command line tool\footnote{\url{https://github.com/github/codeql}} that supports Python and JavaScript and comes equipped with a security suite capable of detecting CWE-079\footnote{\url{https://github.com/github/codeql/tree/main/javascript/ql/src/Security/CWE-079}} and CWE-020 vulnerabilities\footnote{\url{https://github.com/github/codeql/tree/main/python/ql/src/Security/CWE-020}}. It analyzes the entire Git repository and performs a scan for the queried vulnerabilities. CodeQL's output is in compliance with the Static Analysis Results Interchange Format (SARIF)\footnote{\url{https://docs.oasis-open.org/sarif/sarif/v2.1.0/cs01/sarif-v2.1.0-cs01.html}}, returning a list of detected vulnerabilities along with additional information such as their severity, precision of the scan, file paths, and locations of the code where the vulnerabilities were found.

We applied CodeQL on each snapshot of every sampled repository, resulting in a time series of code scanning results for each repository. To obtain a time series with equally spaced points in time, we interpolated missing values. This enabled us to aggregate the repositories' time series into a single time series by calculating the average number of detected vulnerabilities per sampling point. Additionally, since our database stored the type of each detected vulnerability, such as Xss or ReflectedXss, we were able to create type-based time series that aggregated results based on a given vulnerability type.

\subsection{Setup}
CodeQL was executed on a high-performance computing cluster that provided 96 nodes with 28 cores each and 56 GB of memory per node. Each core had two hyperthreads and a nominal frequency of 2.6 GHz. The cluster was shared with other users and managed by an independent work load manager. Code scanning jobs were performed in batches of 250 jobs and processed using a serial queue. Thanks to the computing cluster's parallel processing capabilities, we were able to complete the scanning of all repositories in a reasonable amount of time.

\section{Counterfactual Prediction}
\label{sec:counter_factual}
This section outlines the methodology we used for estimating the causal effect of all three interventions.

We perform counterfactual prediction on the collected time series in order to predict whether the deployed interventions had a significant causal effect on the measured metrics. This type of non-experimental approach to causal inference is required since the interventions have already been deployed and we cannot perform a randomized experiment where a subset of users were not confronted with the intervention.

Counterfactual prediction is a technique that enables us to estimate the causal effect of an intervention by predicting what would have happened in the absence of the intervention. To do this, we use a Bayesian structural time-series model that relies on two types of time series: a response time series and one or more control time series~\cite{brodersen2015inferring}.\footnote{Please refer to the research paper by Brodersen et al.~\cite{brodersen2015inferring} to find more details about the architecture and evaluation of the model.}

The response time series is the data that we collected and described in the previous Section~\ref{sec:time-series-data}. It represents the outcome that we are interested in, such as the updates, revokes and vulnerable commits before and after the intervention.

The control time series, on the other hand, are covariates that we use to help explain the variation in the response time series. For instance, we might use the weather conditions, day of the week, or other relevant factors that could impact the response time series. 

The Bayesian structural time-series model is trained on the response time series from the pre-intervention period, as well as the control time series. The model learns how to explain the response time series as a function of the control time series.

Once the model is trained, we can use it to predict what would have happened in the post-intervention period if the intervention had never occurred. We call this prediction the ``synthetic control'', as it represents a hypothetical scenario where the intervention was not implemented. To make this prediction, we input the control time series from the post-intervention period into the model. 

The difference between the synthetic control and the actual response time series in the post-intervention period gives us an estimate of the causal effect of the intervention.


\begin{figure*}[t]
\centering
    \begin{subfigure}[t]{0.5\textwidth}
        \includegraphics[width=1.0\textwidth]{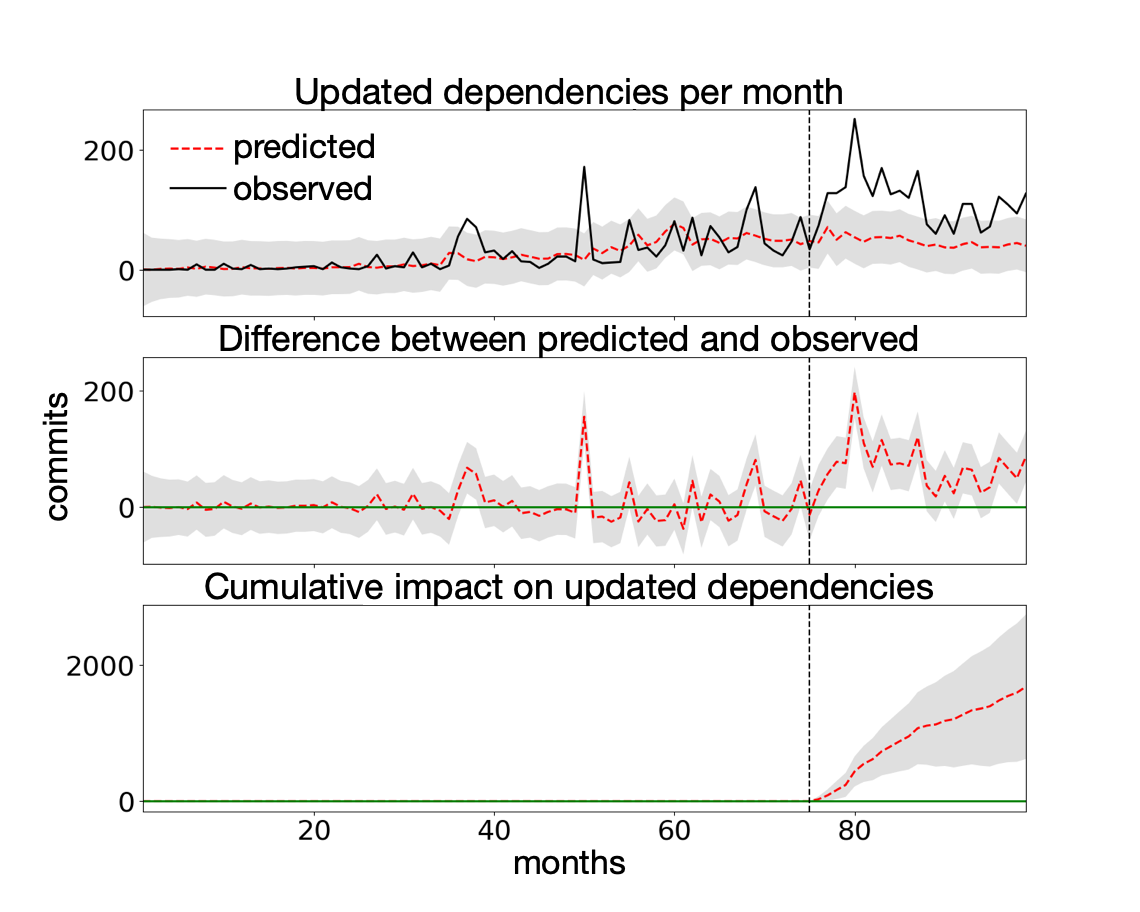}
        \caption{Security alert intervention}
    \end{subfigure}%
    \begin{subfigure}[t]{0.5\textwidth}
        \includegraphics[width=1.0\textwidth]{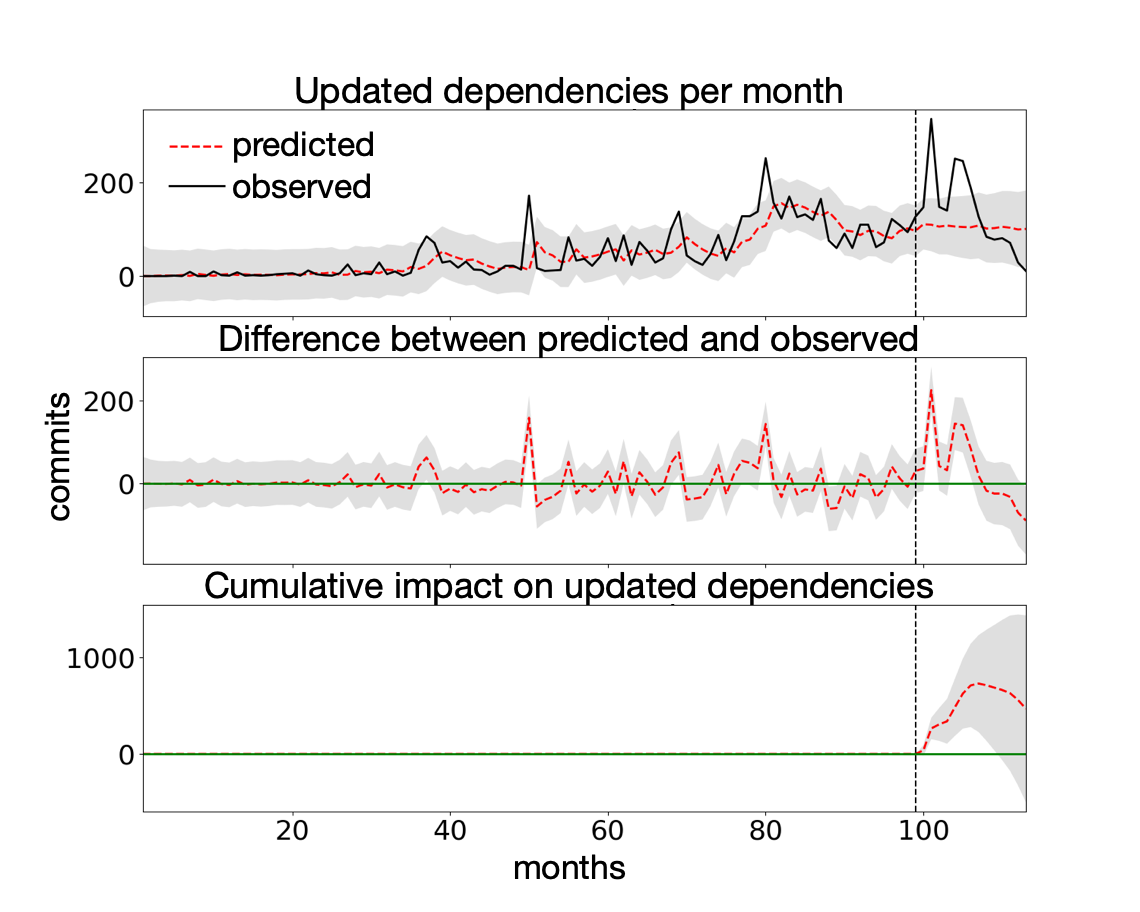}
        \caption{Security update intervention}
    \end{subfigure}
\caption{\diff{Observed and predicted time series of updated vulnerable dependencies per month. The vertical dashed line indicates the day of the deployment of the intervention. The left plot shows the effect of the alert intervention and the right plot the additional update intervention.}}
\label{fig:results_updates}
\end{figure*}

\subsection{Control time series}
To create the control time series \diff{for the security alert and update intervention}, we focused on dependencies that were not present in GitHub's vulnerability list and have not been known to be vulnerable. This time series comprises only those commits that altered such non-vulnerable dependencies. We chose this control time series because we expected it to exhibit some correlation with the three observed time series in the pre-intervention period. By selecting dependencies that are not listed as vulnerable, we aim to capture the underlying patterns that are not related to security issues. This approach helps us to better isolate the causal effect of the interventions on the response time series. \diff{The control time series was created from a set of 40,000 repositories.}

\diff{The control time series for the code scanning intervention comprised 10,000 repositories utilizing JavaScript or Python without CodeQL activation. It tracked the number of commits per time interval that modified Python or JavaScript files.}

\subsection{Necessary condition}
To ensure that the control time series used in our analysis were appropriately matched with the response time series during the pre-intervention period, we conducted correlation tests on both sets of time series. 
We detrended and conducted Pearson tests on all pairs of response and control time series, and found that they all exhibited statistically significant correlations ($p<0.05$). \diff{This established their suitability for the causal inference framework}.



\subsection{Assumptions}
Non-experimental causal inference requires strong assumptions for the control time series in order to be able to draw valid conclusions~\cite{bertrand2002much}. In particular, the following assumptions must hold:

\textit{\textbf{Assumption~1}}: \textit{The control time series is not affected by the intervention.}

This is important because if the control time series is influenced by the intervention, we might mistakenly conclude the existence of an effect or we might under- or overestimate the true effect.

Fortunately, our control time series is unlikely to be affected by the intervention. This is because the intervention only targets dependencies that are present in the vulnerability list, which should not have any impact on the number of commits for dependencies that are not in the list. 



\textit{\textbf{Assumption~2}}: \textit{The relationship between the control time series and the response time series established during the pre-intervention period remains stable throughout the post-intervention period.}

If the correlation between the two time series changes in the post-period, the predicted synthetic control time series may not be accurate.

However, both the control time series and the response time series are measuring the same underlying phenomenon, namely, changes in dependency updates over time. Therefore, it is logical to assume that any relationship established between the two time series during the pre-intervention period would remain stable in the post-intervention period.

We applied the Python package \texttt{causal\_impact} for the analysis and the causal impact visualizations.\footnote{\url{https://github.com/tcassou/causal_impact}}

\section{Results}
\label{sec:results}
\subsection{Security alerts and updates}
\begin{figure*}[t]
\centering
    \begin{subfigure}[t]{0.5\textwidth}
        \includegraphics[width=1.0\textwidth]{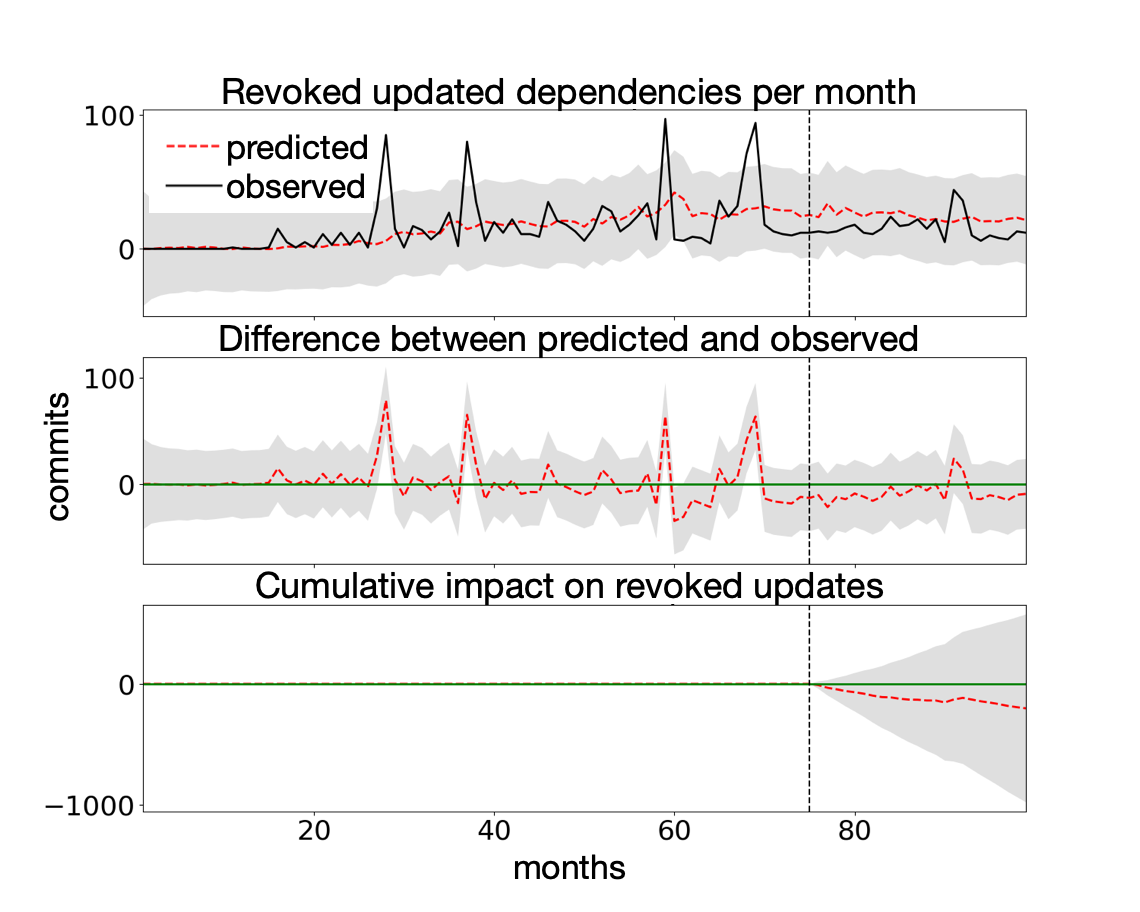} 
        \caption{Security alert intervention}
    \end{subfigure}%
    \begin{subfigure}[t]{0.5\textwidth}
        \includegraphics[width=1.0\textwidth]{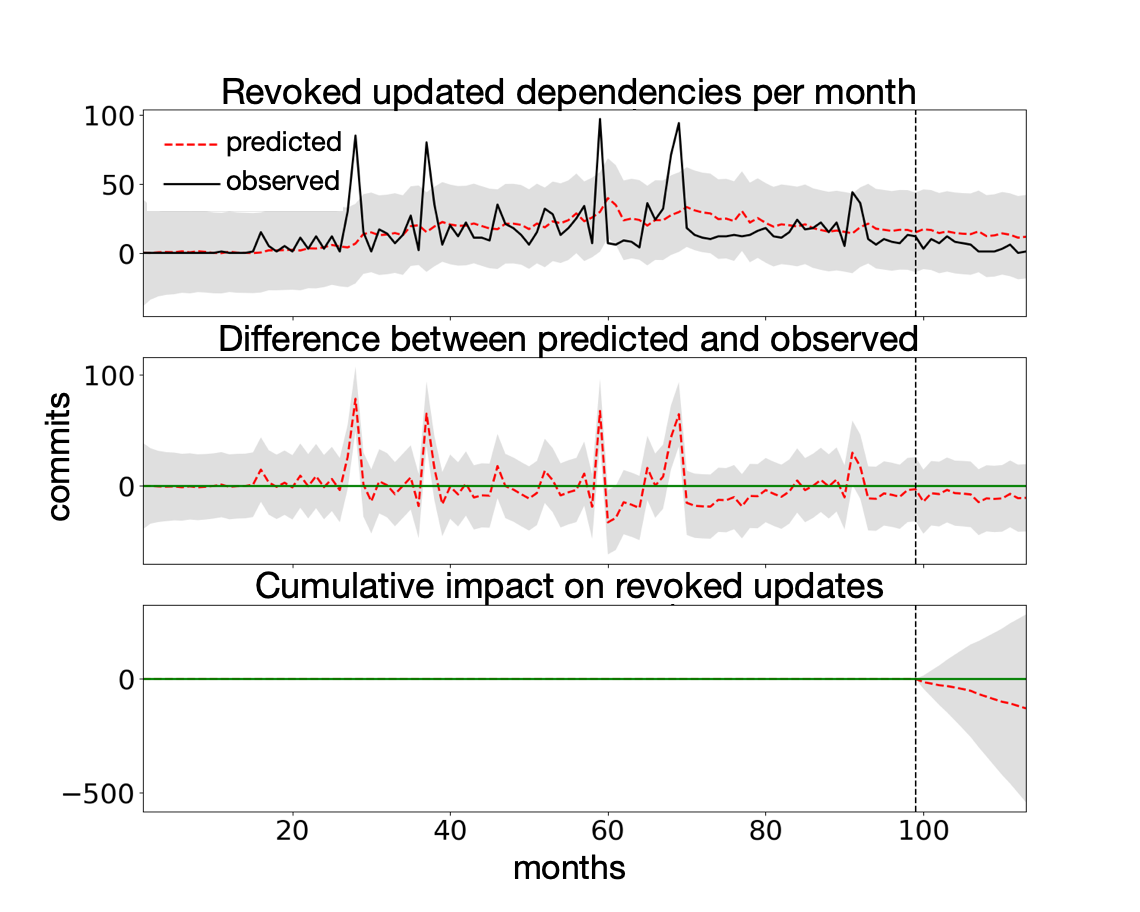}
        \caption{Security update intervention}
    \end{subfigure}
\caption{\diff{Observed and predicted time series of revoked updates per month. The vertical dashed line indicates the day of the deployment of the intervention. The left plot shows the effect of the alert intervention and the right plot the additional update intervention.}}
\label{fig:results_revoked}
\end{figure*}

\begin{figure*}[t]
\centering
    \begin{subfigure}[t]{0.5\textwidth}
        \includegraphics[width=1.0\textwidth]{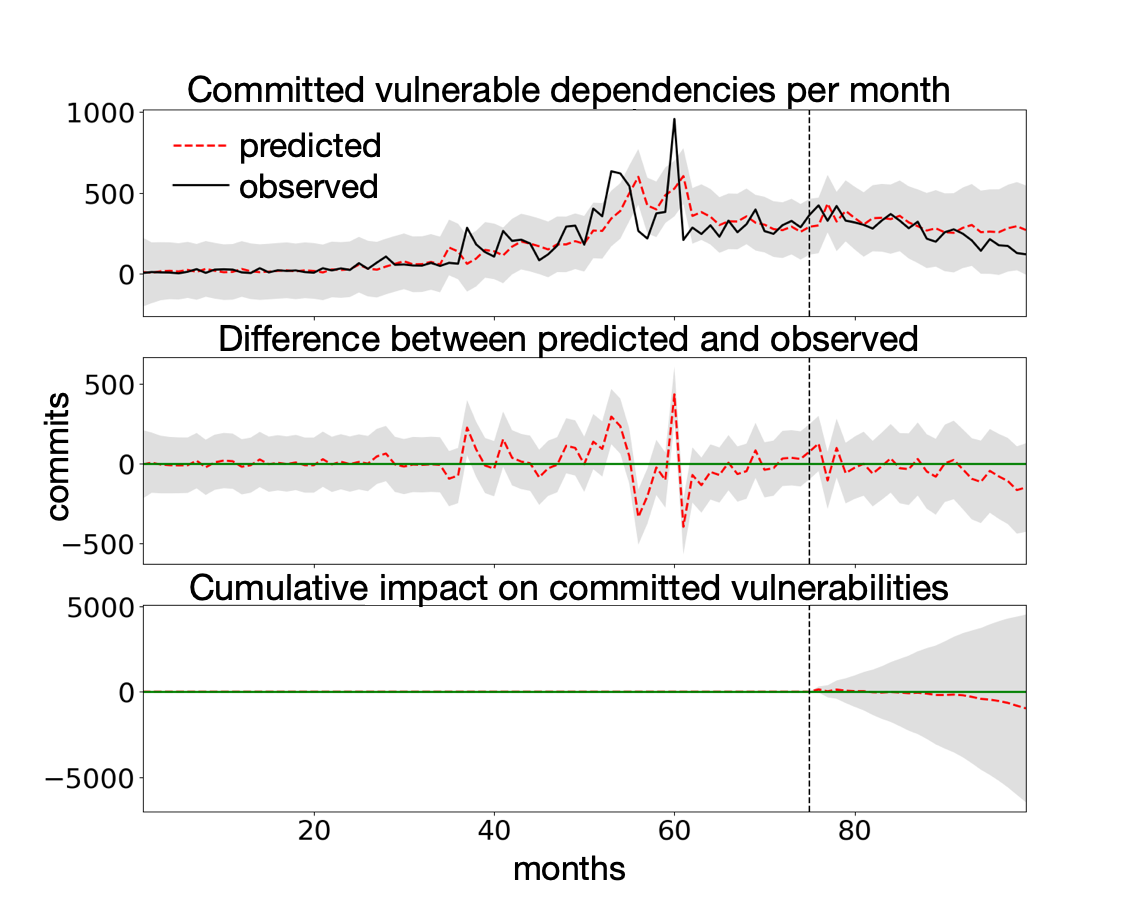}
        \caption{Security alert intervention}
    \end{subfigure}%
    \begin{subfigure}[t]{0.5\textwidth}
        \includegraphics[width=1.0\textwidth]{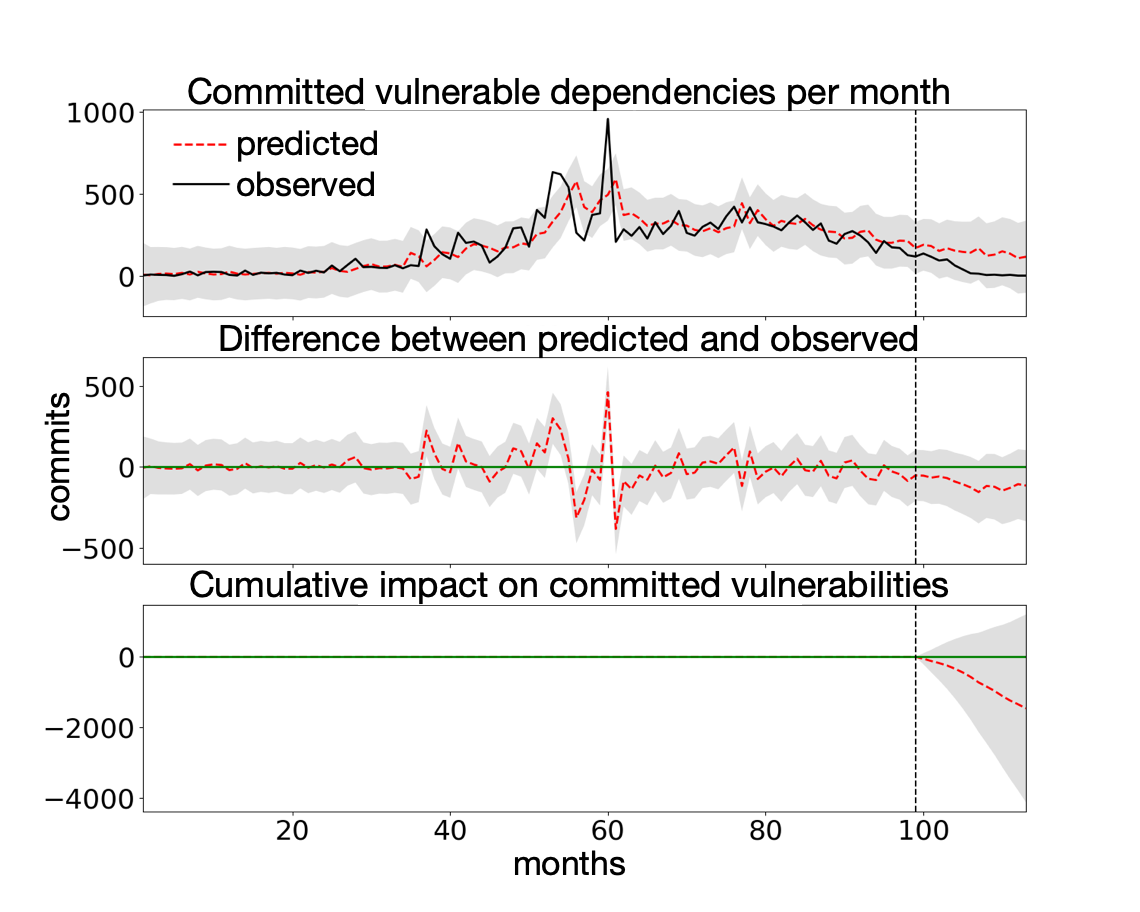}
        \caption{Security update intervention}
    \end{subfigure}
\caption{\diff{Observed and predicted time series of committed vulnerable dependencies per month. The vertical dashed line indicates the day of the deployment of the intervention. The left plot shows the effect of the alert intervention and the right plot the additional update intervention.}}
\label{fig:results_detected}
\end{figure*}

The purpose of this section is to present the results of the causal inference analysis conducted to investigate the impact of security alerts and updates on the use of vulnerable dependencies in GitHub projects. To this end, we collected time series data on updated vulnerable dependencies, revoked updates, and committed vulnerable dependencies, which are visualized in Figures~\ref{fig:results_updates} through \ref{fig:results_detected}. These plots illustrate the response time series and the time series predicted by the model, as well as their difference, and highlight the cumulative impact of the interventions. The results for the security alert and update interventions are shown in the left and right subplots, respectively. In each subplot, the response time series for updates, revokes, and vulnerable commits are represented by a black line labeled `endog'. The predicted time series generated by the model is illustrated by the dashed red line labeled `model'. To facilitate analysis, we have separated both time series into two periods: pre-intervention and post-intervention, which are indicated by a vertical dashed line.




\paragraph{\textbf{Updated vulnerable dependencies pre security alerts}}
Figure~\ref{fig:results_updates} depicts the time series of updated vulnerable dependencies, with the left plot representing the security alert intervention and the right plot showing the additional effect of the security update intervention.

Before the introduction of the security alerts, the upper-left plot (observation vs. prediction) shows that updates were performed rarely, with the daily number of updates hovering around zero to fifty per day. This is illustrated by the response time series, displayed as the black line labeled `endog´ in the upper-left plot. The response time series shows a consistent, slight upward trend with occasional spikes, which flattens out until the introduction of the intervention, indicated by the dashed vertical line. The marginal upward trend observed in the pre-period of the intervention can be attributed to the natural growth in the number of repositories over time, which can result in an increase in any type of commits. The model accurately captures the trend of the response time series during the pre-period, as shown by the red dashed line labeled `model' up until the day the intervention was introduced.


\paragraph{\textbf{Updated vulnerable dependencies post security alerts}}
After the introduction of the security alerts, the response time series shows a marked increase in the number of committed updates per day. Overall, we observed a total of 2,806 updates during the post-intervention period. However, the model predicted that the daily updates would remain relatively constant and flat throughout the entire post-period in the absence of the alert intervention, similar to the pre-period. Specifically, the model predicted only 1,126 updates, with a 95\% confidence interval of $[66, 2186]$. By subtracting this prediction from the observed number of updates, we estimated the causal effect of the alerts on the total amount of updates. The effect was substantial, leading to a 149.0\% increase in the amount of updates, with a 95\% confidence interval of $[243.1\%, 55.0\%]$. The probability of obtaining such an effect by chance is very low, with a Bayesian one-sided tail-area probability of $p=0.019$, indicating that the causal effect is statistically significant. These findings suggest that the alerts had a strong positive impact on the amount of updates made, resulting in a significant increase in the overall level of security.

To recap, we found a surprisingly large effect of security alerts with an increase of 149\% updates on average. While prior work has shown that security alerts are generally effective, they tend to have a relatively small effect size. The positive effect could be caused by the thorough alert design GitHub has applied. First of all, GitHub offers different notification types for security alerts. Users are notified by email, command line interface (on push events), and on GitHub by default. Moreover, alerts include CVSS metrics, which allow users to better evaluate the risk. For instance, it allows users to understand that confidential data can be stolen remotely by performing a relatively easy attack that does not require any privileges or user interaction. Finally, GitHub's security alerts provide actionable recommendations on how to fix detected issues. These include a list of patched versions of the vulnerable dependencies. Previous work has shown that informing users about potential risks and attack vectors, while simultaneously recommending actions that provide a way out of the warning situation, improve the effectiveness of warnings~\cite{reeder2018experience}. However, users still had to find and install the recommended versions themselves. This is an additional burden, which might have contributed to users not performing an update. Unfortunately, GitHub does not provide a recommended action in case a patched version is unavailable. 

\paragraph{\textbf{Updated vulnerable dependencies pre security updates}}
In Figure~\ref{fig:results_updates}, we present the plots on the right side that depict the impact of the update intervention that was implemented at a later stage. The upper-right plot displays the response time series of updates represented by the black line labeled `endog', while the predicted time series generated by the model is illustrated by the dashed red line labeled `model'. Both time series were separated into two periods: pre- and post-intervention, indicated by a vertical dashed line.

Prior to the update intervention, the impact of the alert intervention can still be observed as its post-period ends just before the update intervention. During this period, we observed a decrease in the effect of the alert intervention, with the number of updates per month showing a slight downward trend. 

\paragraph{\textbf{Updated vulnerable dependencies post security updates}}
Immediately after the update was deployed, the number of updates per month showed a sharp increase, averaging 138 updates per month. Interestingly, our model predicted that the number of updates would continue to stagnate, estimating an average of 105 updates per month in the absence of intervention (95\% CI: $[36, 174]$). This represents an increase of updates of 31.7\% (95\% CI: $[97.3\%, -33.9\%]$). The causal effect is statistically significant (Bayesian one-sided tail-area probability $p=0.001$).

Based on our analysis, it appears that the security alerts intervention was more effective, resulting in a 149\% increase in updates per month. However, it is important to consider that the update intervention was built on top of the security alerts and implemented much later in time. This means that the impact of the security alerts was already present in the data when the update intervention was introduced. The model accounts for this by predicting that the number of updates per month would be 105 if the update intervention had not occurred, which is similar to the amount observed during the post-period of the security alerts intervention. In conclusion, the update intervention improved upon the security alerts intervention by an additional 31.7\%.

We also need to take into account the relatively shorter post-period of the update intervention. Therefore, the effect of the intervention could become much more significant over time. As shown in the bottom-right plot in Figure~\ref{fig:results_updates}, the cumulative impact of the update intervention has a steeper increase compared to security alerts as shown in the bottom-left plot. This indicates that more updates occurred during the first few months after the update intervention was implemented. These additional improvements of the update intervention are reasonable as automating security updates reduces the burden on users. Given the higher frequency of updates during the post-period of the update intervention, we expect it to become even more effective than security alerts.




\paragraph{\textbf{Revoked updates pre and post security alerts}}
The upper-left plot in Figure~\ref{fig:results_revoked} shows a very subtle upward trend in the amount of revokes per month in the pre-period of the security alert intervention. This trend can be explained by the increasing amount of repositories that were created on GitHub. 

However, during the post-period of the security alert intervention, we observed a downward trend in the average number of revoked updates. On average, 15 updates were revoked each month, which is lower than the expected 24 updates that would have been revoked in the absence of the intervention (95\% CI: $[-8, 56]$). This corresponds to a 34.8\% reduction in revoked updates (95\% CI: $[98.6\%, -168.1\%]$). The causal effect is not statistically significant (Bayesian one-sided tail-area probability $p=0.071$).

Considering the effect of the update intervention shown in the right plot, we see a slightly stronger decrease in the amount of revokes in the post-period. Here, only four updates were revoked each month on average, while the model predicted an average of 13 revokes if no intervention happened (95\% CI: $[-15, 43]$). This is a decrease of 66.3\%. However, the causal effect is again not statistically significant (Bayesian one-sided tail-area probability $p=0.177$).

While the interventions aimed to increase security updates, the effectiveness would be undermined if a significant number of updates were unusable and revoked. This could be due to updates that introduce issues such as broken code or compilation errors.

Surprisingly, we found that both interventions, security alerts and updates, had similarly high levels of usability, as evidenced by a reduction in revokes during the post-intervention period. This finding was unexpected, as we had anticipated a potential increase in revokes due to the higher volume of updates generated by the interventions. Nevertheless, the effect was not statistically significant. Note that a significant decrease in revokes was not necessarily required for the interventions to be considered successful since the pre-intervention rate of revokes was already relatively low. However, the absence of an increase in revokes is crucial to the success of the interventions, as any increase may lead users to disable the interventions entirely.

\paragraph{\textbf{Vulnerable dependencies committed pre security alerts}}
The upper-left plot in Figure~\ref{fig:results_detected} depicts an exponential increase in the number of committed vulnerable dependencies, as indicated by the black line labelled `endog', from January 2014 to March 2016. This trend is attributed to the exponential growth of GitHub in the size of newly created repositories. However, immediately following the deployment of the security alert intervention, we observe a shift in the pattern towards a gradual decrease. The vertical dashed line in the plot indicates the timing of the security alert intervention. The red dashed line represents the prediction of the Bayesian time series model. The predicted time series in the post-period represents the synthetic control. 

\paragraph{\textbf{Vulnerable dependencies committed post security alerts}}
During the post-period, an average of 264 vulnerable dependencies were committed per month, while the counterfactual prediction estimated a value of 304 vulnerable dependencies per month in the absence of the alert intervention (95\% CI: [76, 533]). This corresponds to a causal effect estimate of around -40 (-13.4\%) on the amount of committed vulnerabilities per month. Although the effect was statistically significant, the effect size was relatively small (Bayesian one-sided tail-area probability $p=0.004$).

The impact of security alerts on educating users to install fewer vulnerable dependencies was found to be small. Despite GitHub's detailed information on detected vulnerabilities, including criticality score, affected and patched version, and labels, users were not likely to remember this information or look it up to avoid installing vulnerable dependencies again. As a result, users relied more on the intervention to notify them when they installed a vulnerable version so they could update it later.

\paragraph{\textbf{Vulnerable dependencies committed post security updates}}
The right plot in Figure~\ref{fig:results_detected} depicts the effect of the update intervention, which was deployed later. We observe a slightly stronger decreasing trend in committed vulnerable dependencies. During the post-period, on average, 42 vulnerable dependencies were committed per month, whereas the counterfactual prediction expected a value of 147 (95\% CI: [-43, 337]). The update intervention caused a 71\% decrease in committed vulnerabilities per month. Although the effect size is considerable, the causal effect was not statistically significant (Bayesian one-sided tail-area probability $p=0.065$).

The larger decrease in the number of committed vulnerable dependencies during the post-period of the update intervention can be explained by overlapping effects with the security alert intervention. It is worth noting that the update intervention runs on top of the security alert intervention. Therefore, the decrease observed in the update intervention may be partially due to the increasing educational effects of the security alert intervention.

\paragraph{\textbf{Variance among severity}}
We analyzed whether the interventions had different effects on vulnerable dependencies of different severity levels, specifically for critical, high, moderate, and low severity. The results showed that both interventions, i.e., security alerts and updates, were particularly effective in reducing vulnerabilities with critical and high severity, while their effect was less pronounced for low severity issues. This suggests that users are more responsive to vulnerabilities with higher severity levels, possibly due to the perceived greater potential impact of such vulnerabilities. It is possible that users rely on the given CVSS criticality scores and metrics to determine the appropriate action to take for each vulnerability.


\subsection{Code Scanning}
CodeQL found at least one vulnerability in 1,842 repositories, i.e., 18.42\% of the examined 10,000 projects. Note that for 1,059 projects CodeQL was not able to perform an analysis due to occurring errors. We excluded them from further analysis. 
We report time series data of detected CWE-079 vulnerabilities in JavaScript and CWE-020 vulnerabilities in Python. While our analysis mainly focused on JavaScript and CWE-079, we additionally investigated Python and CWE-020 in order to investigate whether we observe different effects depending on programming language and vulnerability class. In total, we analyzed 437,485 repository snapshots for the CWE-079 detection in JavaScript and 54,040 snapshots for CWE-020 detection in Python. 

\begin{figure*}[t]
\centering
    \begin{subfigure}[t]{0.5\textwidth}
        \includegraphics[width=1.0\textwidth]{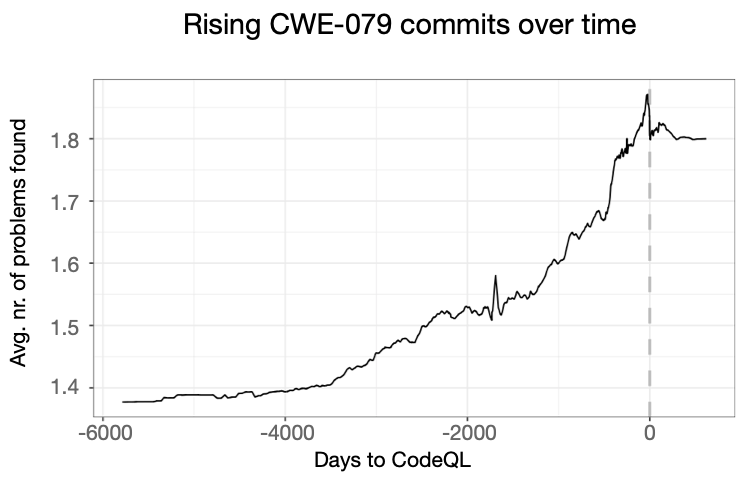} 
    \end{subfigure}%
    \begin{subfigure}[t]{0.5\textwidth}
        \includegraphics[width=1.0\textwidth]{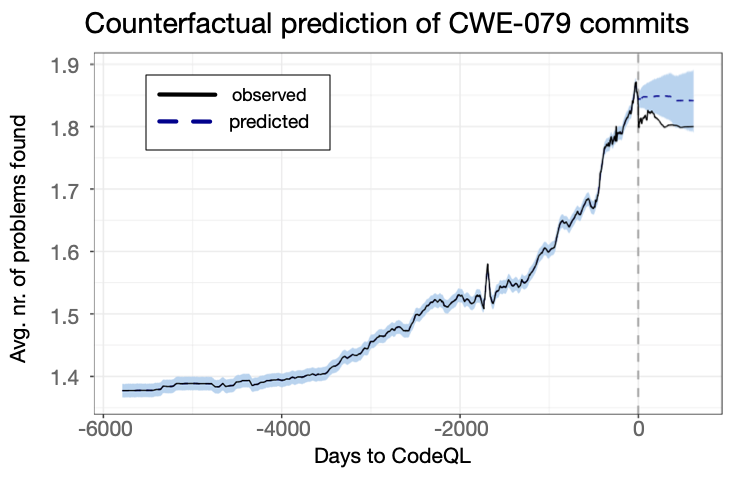}
    \end{subfigure}
    \begin{subfigure}[t]{0.5\textwidth}
        \includegraphics[width=1.0\textwidth]{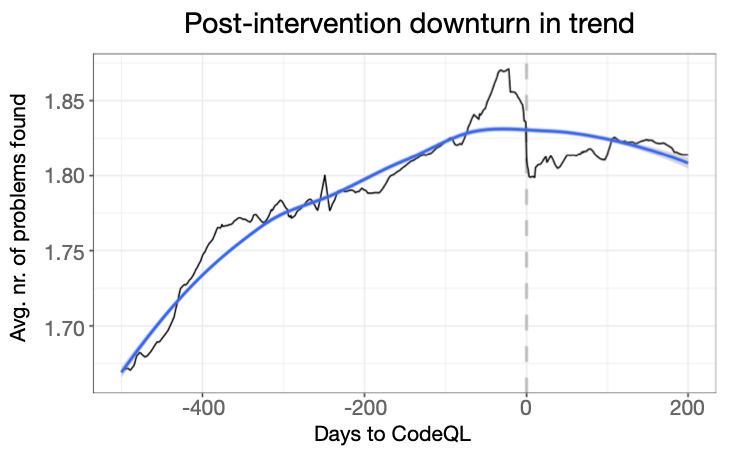} 
    \end{subfigure}%
    \begin{subfigure}[t]{0.5\textwidth}
        \includegraphics[width=1.0\textwidth]{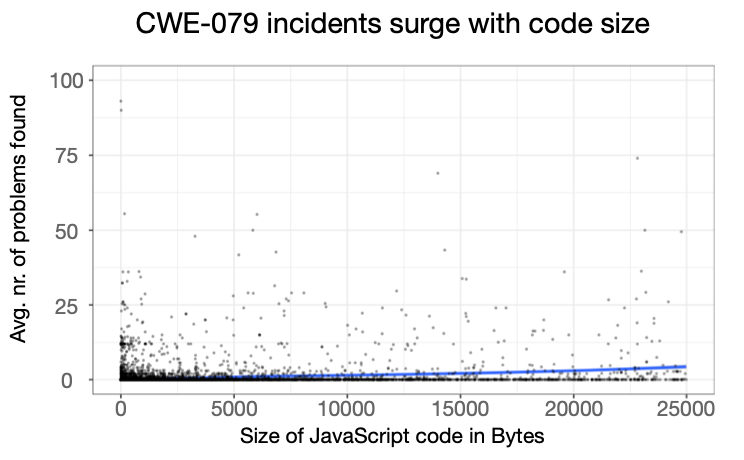}
    \end{subfigure}%
\caption{\diff{The upper-left plot shows the time series of the daily commited average amount of CWE-079. The dashed vertical line indicates activation of CodeQL. The dashed blue line in the upper-right plot shows the counter factual prediction of the time series in the post-period. The lower-left plot shows the local regression of the time series, while the lower-right plot shows the relation between detected CWE-079 and amount of JavaScript code indicated by the blue line.}}
\label{fig:overall_results}
\end{figure*}

\begin{figure*}[t]
\centering
    \begin{subfigure}[t]{0.5\textwidth}
        \includegraphics[width=1.0\textwidth]{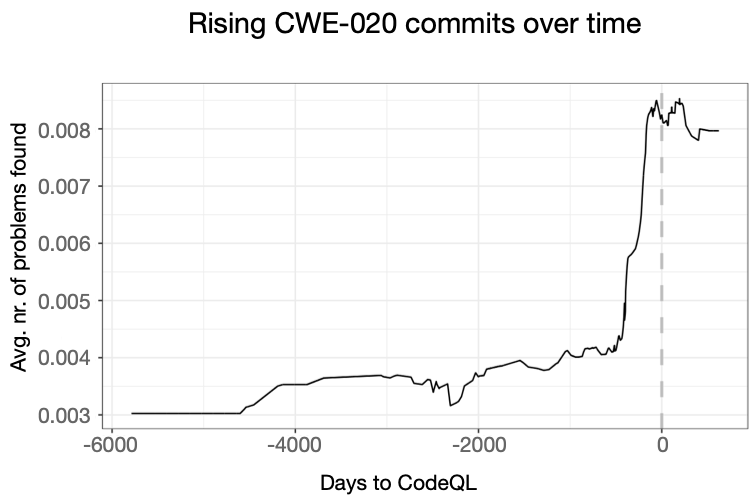} 
    \end{subfigure}%
    \begin{subfigure}[t]{0.5\textwidth}
        \includegraphics[width=1.0\textwidth]{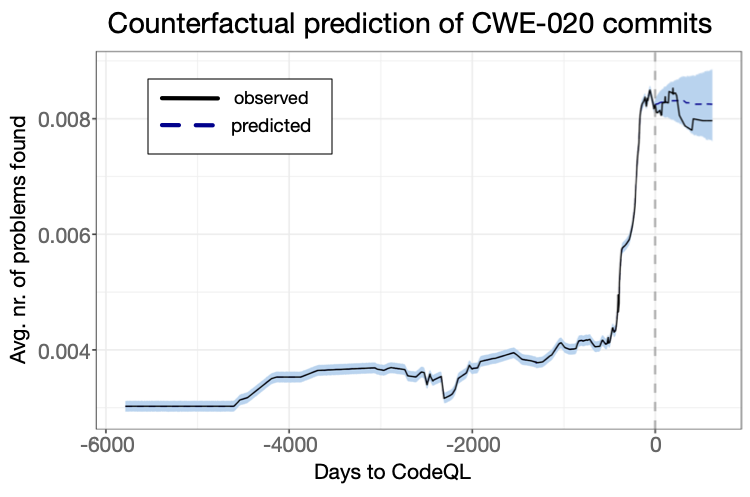}
    \end{subfigure}
    \begin{subfigure}[t]{0.5\textwidth}
        \includegraphics[width=1.0\textwidth]{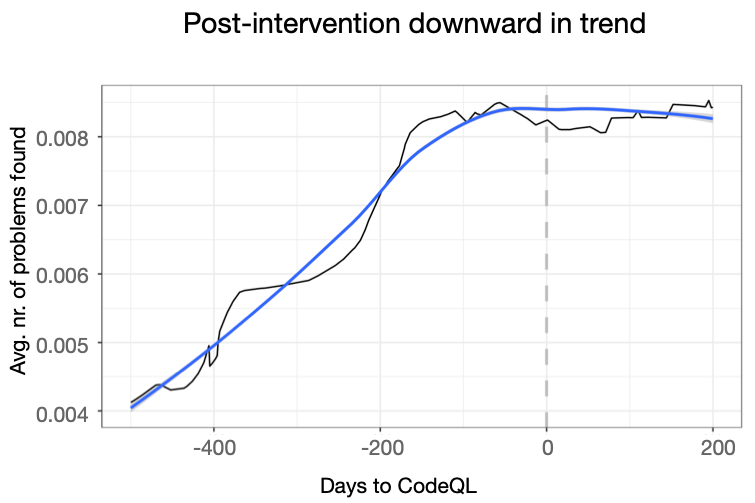} 
    \end{subfigure}%
    \begin{subfigure}[t]{0.5\textwidth}
        \includegraphics[width=1.0\textwidth]{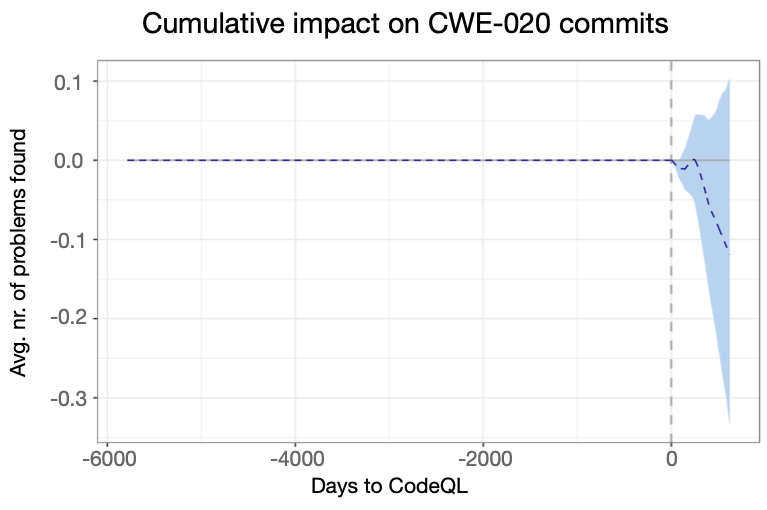}
    \end{subfigure}%
\caption{\diff{The upper-left plot shows the time series of the daily commited average amount of CWE-020. The dashed vertical line indicates activation of CodeQL. The dashed blue line in the upper-right plot shows the counter factual prediction of the time series in the post-period. The lower-left plot shows the local regression of the time series, while the lower-right plot shows the cumulative impact of the code scanning intervention on the mean of commited CWE-020 per day and its confidence interval.}}
\label{fig:overall_results_cwe020}
\end{figure*}

\paragraph{\textbf{Vulnerability accumulation \& decline}}
The upper left plots in Figure~\ref{fig:overall_results} and Figure~\ref{fig:overall_results_cwe020} show the time series as introduced in Section~\ref{sec:time-series-data}. They show the average number of vulnerabilities found in a repository on the y-axis and the days to activation of CodeQL in the x-axis. Day zero---marked by the dashed grey line---indicates the day where each repository activated CodeQL. The plot shows that repositories accumulate vulnerabilities over time. We assumed that the amount of vulnerabilities increased gradually because of the increasing size of the code base of these repositories. The local regression in the lower right plot in Figure~\ref{fig:overall_results} (the blue line indicates the moving average) shows that the average number of problems found indeed increases with the size of the code base. However, the large scatter also means that this effect should not be overstated. 

Since the activation of CodeQL, the average number of CWE-079 and CWE-020 has decreased as shown in the upper right plots in Figure~\ref{fig:overall_results} and Figure~\ref{fig:overall_results_cwe020}. The dashed blue line indicates how the model would have expected the time series to look like, if the intervention would not have happened. The steep decline of vulnerabilities right after the intervention suggests a novelty effect. Users might have had an increased motivation to follow CodeQL warnings and to fix the code on account of the newness of the intervention. Both time series have a small spike right after the drop and then flatten out having an overall negative trend as shown by the local regression in the lower left plot in Figure~\ref{fig:overall_results} and Figure~\ref{fig:overall_results_cwe020}. We applied the Python package \texttt{causal\_impact} to analyze whether this downward trend was caused by the code scanning intervention and whether the effect was significant.

\paragraph{\textbf{Seasonal effects}}
\diff{Figures \ref{fig:overall_results} and \ref{fig:overall_results_cwe020} depict notable fluctuations in the frequency of detected problems for both CWE types. Following an initial decline immediately after the intervention, a subsequent trend emerges wherein the count of detected problems experiences a resurgence after a span of several days, only to exhibit a subsequent decline once more.

One possible explanation for these fluctuations is that CodeQL predominantly operated on a fixed-schedule plan, leading to the accumulation of vulnerabilities over time before they were identified and addressed. Consequently, following the initial decrease, there was an increase in the number of detected problems after the intervention. These seasonal effects (periodical fluctuations in the number of detected problems) account for the stagnant trend evident in both figures. However, it is important to note that the overall trend after the intervention exhibits a negative trajectory.

We propose that the identification of seasonal effects (periodical fluctuations in the number of detected problems) serves as a supporting factor indicating that CodeQL is the causal intervention. This assertion is strengthened by the fact that CodeQL is implemented on a fixed schedule.}

\paragraph{\textbf{Causal effect on CWE-079}}The upper right plot in Figure ~\ref{fig:overall_results} covers the time period from 500 days prior to activation and 200 days after activation of CodeQL. The black line shows the measured time series of detected vulnerabilities, while the blue dashed line shows the time series---the synthetic control---predicted by the model. The area highlighted in blue shows the 95\% confidence interval. During the post-period, CodeQL detected an average of 1.81 vulnerabilities. Without the intervention, however, we would have expected an average value of 1.85, (95\% CI: $[1.82, 1.87]$). This means that maintainers of a repository fixed an average of 0.039 vulnerabilities (95\% CI: $[0.011, 0.067]$) that CodeQL alerted them about. This corresponds to a 2\% reduction in vulnerabilities, with a 95\% confidence interval of $[1\%, 4\%]$. The posterior probability of a causal effect of CodeQL is 99.2\%. The Bayesian one-sided tail-area probability of observing this effect by chance is $p = 0.008$.

When creating a new repository the code scanning intervention is not active per default. We have identified a rather small amount of repositories where the intervention was activated by the user. For these repositories, the effect size was moderate with a reduction of 2\% of detected vulnerabilities. A large fraction of users did not seem to have paid attention to the warnings or might have missed them. In response, we investigated the communication design of detected problems. We found that there might be issues with the design of the security indicators; they are very small and not easily visible in the main menu of the repository\ifthenelse{\boolean{isArxiv}}{ (see left screenshot in Figure~\ref{fig:codeql_ui}).}{ (please find screenshots of the UI and user journey in the appendix of the author version of this paper).\footnote{\url{https://arxiv.org/abs/2309.04833}}} Moreover, users have to follow a long path in order to find the specific results of the code scan\ifthenelse{\boolean{isArxiv}}{ (see right screenshot in Figure~\ref{fig:codeql_ui}).}{.} Even though they also provide recommendations on how to fix the identified issues, these only appear after clicking an expand button whose description does not suggest the availability of recommendations. In contrast to the update intervention for vulnerable dependencies the recommendations do not provide quick fixes.

We found that most users that patched vulnerabilities found
by CodeQL tend to patch more than 95\% of the problems. We
assume these users to be very security-aware and therefore to
have higher willingness to pay attention to the CodeQL security indicators, findings and recommendations than average
users. 

\paragraph{\textbf{Variance among CWE-079 types}} The largest observed significant effect of CodeQL was on fixing DOM Text Reinterpreted as HTML (\textit{XSS through DOM}). The average amount of vulnerabilities decreased by three per cent (95\% interval: $[2\%, 5\%]$, $p < 0.001$). However, the downward trend right after the intervention seems to stagnate. This suggests that an alert about a security vulnerability that is not immediately fixed is likely to be ignored even in the longer term. At the same time, however, the stagnation also contrasts with the previous observation that problems accumulate over time. Thus, CodeQL presumably has educational effects that prevent problems from being added to the code base in the first place. Interestingly, 44\% of the repositories that followed the warnings tend to do this thoroughly as they removed 95\% of the problems. However, a large fraction of repositories (71.7\%) seemed to ignore the warnings or missed them as they did not fix any of the detected problems.

We saw a similar significant effect on \textit{Unsafe jQuery Plugin} as there is a decrease of about 4\% in identified problems (95\% interval: $[2\%, 6\%]$, $p=<0.01$). 78.9\% of repositories did not patch any of the reported problems, while about 5\% removed 95\% of the issues and 5.5\% at least improved some of the code. However, we also identified 10.6\% of the repositories where performance decreased. These repositories had more vulnerabilities after the intervention than before.

We identified a significant effect on \textit{Reflected XSS} very similar to \textit{Unsafe jQuery Plugin}---but with a bigger effect size---as the average amount of problems was reduced by about 6\%.

We saw counter-intuitive results for \textit{Unsafe HTML Constructed From Library Input, Client-Side Cross-Site Scripting}, and \textit{XSS through Exception}. Here, the intervention had a significant negative causal effect on the reduction of detected problems. For \textit{Client-Side Cross-Site Scripting}, we found two outliers in the sample that were responsible for this result. One repository committed 24 vulnerabilities, while the other one committed 127 vulnerabilities within a single day right after activating CodelQL. After manually checking the committed vulnerabilities, we found that those were introduced on purpose to test CodeQL's detection accuracy. As correctly predicted by \texttt{causal\_impact}, the intervention caused those repositories to add more vulnerabilities.

\textit{Stored XSS} was only present in 11 repositories. Therefore, the sample was too small to perform any meaningful statistical analysis. However, we identified one repository that patched all 20 identified vulnerabilities, which had accumulated over a period of three years shortly after the repository activated CodeQL.


\paragraph{\textbf{Causal effect on CWE-020}}
During the post-intervention period, we observe a decrease of 2.3\% in the average amount of vulnerabilities per repository. The 95\% interval of this percentage is [-6\%, +2\%]. The probability of obtaining this effect by chance is $p$ = 0.144. This means the effect may be spurious and would generally not be considered statistically significant.

Even though the intervention appears to have caused a positive effect, this effect is not statistically significant when considering the entire post-intervention period as a whole. Individual days or shorter stretches within the intervention period may of course still have had a significant effect, as indicated whenever the upper limit of the confidence interval of the impact time series (lower-right plot in Figure~\ref{fig:overall_results_cwe020}) was below zero. The apparent effect could be the result of random fluctuations that are unrelated to the intervention. This is often the case when the intervention period is very long and includes much of the time when the effect has already worn off. It can also be the case when the intervention period is too short to distinguish the signal from the noise. The statistical insignificance of the intervention could also be explained by the very small amount of detected vulnerabilities per repository.

\paragraph{\textbf{Variance among CWE-020 types}} 
CWE-020 contains different vulnerability types, i.\,e., \textit{ Incomplete regular expression for hostnames, Incomplete URL substring sanitization}, and \textit{Overly permissive regular expression range}. We observed different effects on each of these vulnerabilities.

We observed a statistical significant impact on \textit{Incomplete regular expression for hostnames} (Bayesian one-sided tail-area probability $p = 0.001$) during the first 200 days of the post-period. The identified causal effect of the intervention led to a decrease of 44\% in the average amount of vulnerabilities per repository. Considering the complete post-period we actually observe an increase of 5\% which is not statistically significant.

The observed fluctuations during the post-period can be explained by novelty effects where the intervention is heavily applied in the beginning of the post-period, while its use declines over time. The time series shows a steep decline right after the introduction of the intervention. However, after 150 days it increased up to pre-intervention level. 

The intervention had a significant causal effect  on \textit{Incomplete URL substring sanitization} within the first 200 days of the post-period (Bayesian one-sided tail-area probability $p = 0.002$), as the average amount per repository decreased by about 18\% (95\% interval is [-27\%, -8\%]). Considering the complete post-period, the effect became a bit stronger with an overall increase of 19\% ($p$ = 0.001).

This vulnerability type was the one where the intervention had the largest significant causal effect in comparison to all other vulnerability types in CWE-079 and CWE-020; perhaps due to the low complexity in patching the vulnerability. For instance, in comparison to \textit{Incomplete regular expression for hostnames}, a developer merely has to write down a whitelist of domains, which can be seen as much easier than fixing a regular expression. Even though we observed fluctuations in the post-period, they were very small and the time series maintained its downward trend. Such fluctuations can be explained by seasonal effects occurring due to scheduled repository scanning. E.g., when using a monthly scan schedule, vulnerabilities can accumulate over time until they are being detected and patched.

We did not observe any significant effects on \textit{Overly permissive regular expression range} even though we see a reduction of 1\% over the whole post-period. This can as well be explained due to the complexity of a patch. It requires the developer to define unambiguous regular expressions, which can be seen as a much harder task than patching \textit{Incomplete URL substring sanitization} or \textit{Incomplete regular expression for hostnames}.



\paragraph{\textbf{Variance among repositories}}

The effectiveness of CodeQL varies greatly across different repositories, and its impact on vulnerability reduction depends on several factors. For instance, out of the sample considered for this study, only 15.9\% of repositories with vulnerabilities were improved by CodeQL, while 67.6\% did not improve and 16.4\% actually worsened. 262 repositories added new security vulnerabilities during the post-activation period. It is worth noting that on average, the repositories improved in reducing vulnerabilities.

The reasons for the varied effectiveness of CodeQL could be due to several factors. For instance, the development team may have been active during the post-activation period, and they could have introduced many changes to the code base that increased the likelihood of new vulnerabilities being added. Additionally, the activation of CodeQL could have given a false sense of security, leading to users not paying enough attention to the security indicators. This could explain the relatively high number of repositories where no change was observed. False positives could also contribute to the habituation effect, where users become used to alerts and do not pay enough attention to future alerts.


\paragraph{\textbf{Variance among scan frequency and triggers:}} In our study, we found that the majority of repositories (8,543) used the default weekly scan frequency, while only a small percentage (398) applied customized frequencies. Interestingly, we observed that repositories using the default configuration had a significant reduction in vulnerabilities after the CodeQL intervention, while those with custom configurations showed large fluctuations and even an increase in problems. However, due to the small sample size of repositories with customized configurations, \texttt{causal\_impact} was not able to identify a significant effect.

It is worth noting that CodeQL scans every push to specified branches by default, but a small percentage (1.7\%) of repositories in our sample decided to change this setting. These repositories had 6\% fewer problems on average in the post-period than repositories using the default setting, but this effect was not statistically significant.

Similarly, only 1.8\% of repositories deviated from the default setting of having pull requests trigger CodeQL scans. These repositories had 5\% fewer problems on average in the post-period than repositories using the default setting, but again, the effect was not statistically significant.

\paragraph{\textbf{Variance among popularities}} 
The popularity of a repository on GitHub can be indicated by the number of stars it has received. In our sample, 7,432 projects had 10 or fewer stars, while 1,509 repositories had more than 10 stars. For repositories with more than 10 stars, we found that the causal effect was not statistically significant ($p=0.59$). However, for projects with 10 or fewer stars, the intervention has led to a significant improvement of 3\% (95\% confidence interval: [1\%, 4\%], $p=0.001$).

\section{Confounders}
\label{sec:confounders}

\diff{
Our causal analysis is based on observational data, which differs from randomized data in that it can be susceptible to confounding due to the presence of unobserved variables and their interactions. This raises concerns about the potential influence of these confounding factors on the measured effects. For instance, other factors such as security training or tools unrelated to GitHub's interventions could have caused the observed effects. Additionally, external sources like blog posts, news sites, or mainstream media that raise awareness of vulnerabilities might also play a role.

In the following discussion, we thoroughly examine the plausibility of these confounding effects and assess their potential impact on our findings.

\paragraph{\textbf{Methodical mitigation}}
In our analysis, we employed a Bayesian model, specifically \texttt{causal\_impact}, to evaluate the incremental impact of the intervention during the post-period. As depicted in Figure ~\ref{fig:results_updates}, the cumulative impact of the security alerts exhibits a consistent and linear upward trend, indicating a sustained and accumulating effect over time. The enduring impact of the intervention suggests that one-time interventions, such as blog posts or news articles, are unlikely to be responsible for the observed cumulative effect, which is in agreement with findings by Fiebig et al. that such interventions lack lasting influence after the intervention period~\cite{fiebig2016one}.

For the code scanning intervention, we utilized a control time series, which naturally helps in identifying confounding factors, to compare with the observed time series. The control time series measured the same variables as the observed time series but did not undergo the code scanning intervention. Hence, confounding events that might have influenced the observed time series would likely be present in the control time series as well. The absence of similar effects in the control time series supports the conclusion that the intervention caused the observed effects.

\paragraph{\textbf{Security training and independent tools}} We consider confounding interventions, such as an increase of security training or use of independent security tools that are not integrated into GitHub, to be less plausible. These confounding factors would have had to occur concurrently with the security interventions on GitHub, impacting the same set of users. The probability of multiple independent events coinciding with each of the three intervention dates diminishes the likelihood of such occurrences.

\paragraph{\textbf{Third-party dependency checkers}}
Snyk exhibits substantial resemblance to GitHub's security update intervention with over 175 thousand installations on GitHub (in August 2023), signifying its widespread adoption nowadays.\footnote{Install counts as provided on GitHub marketplace: \url{https://github.com/marketplace}.} However, its introduction occurred only approximately six months after the release of GitHub's intervention. Consequently, the possibility arises that Snyk's presence could have influenced the results toward the end of the six month time window after the introduction of GitHub's intervention. Notably, the most substantial impact of GitHub's intervention was observed in the initial months when Snyk had not yet become available.

\paragraph{\textbf{Third-party code scanners}}
Offering an alternative approach to CodeQL, GitHub allows integration of third-party code scanners into workflows, with a selection of 70 diverse scanners currently available (in August 2023). We assessed the prevalence of third-party code scanner integration within the 37,856 repositories that used CodeQL, by systematically reviewing the repositories' workflow directories, specifically targeting YAML files indicative of third-party scanner adoption. A subset of 168 repositories (0.4\%) was identified as having chosen to employ supplementary third-party scanners alongside their utilization of CodeQL. Given the relatively modest frequency of this co-adoption, it is reasonable to infer that the incorporation of these additional tools did not significantly confound the overall research outcome.




\section{Limitations}
\label{sec:limitations}
In contrast to the security alert intervention, the default activation of security updates and code scanning was not in effect. This disparity could potentially introduce a bias in the sampled repositories, favoring those whose users exhibit a higher degree of concern for security matters.

Our study focused on code scanning within repositories utilizing scheduled scan plans, possibly omitting users who exclusively initiated CodeQL scans through push or pull events.

Furthermore, there exists a possibility of interaction effects between updates for dependencies that are not vulnerable and updates triggered specifically for vulnerable dependencies. In scenarios where users are prompted to update certain dependencies within their repositories, a tendency might arise to update all dependencies simultaneously. Moreover, the security updates for some dependencies could necessitate concurrent updates to other interconnected libraries. However, should we acknowledge the presence of these interaction effects within the control time series, it follows that our analysis might have actually underestimated the true effect, rather than overestimating it.


}

\section{Discussion}
\label{sec:discussion}

Based on the results of our study, we can offer suggestions for tool and intervention creators, as well as educators, to take advantage of GitHub's approach. Additionally, we will provide insights into how GitHub could further improve the design of their interventions.

\paragraph{\textbf{Default interventions}}
Enabling the security alert intervention by default appears to have played a significant role in the effectiveness of the interventions. As previous research in both behavioral science and code security has demonstrated, safe defaults can be highly effective~\cite{fischer2021effect}. This suggests that applying the principle of safe defaults to the code scanning intervention could further improve its impact. 

\paragraph{\textbf{Seamless interventions}} 
The interventions were seamlessly integrated into the software development process and Git environment to minimize disturbance to users. Security alerts were integrated into developer tools such as the command line interface, and the update intervention was automated, allowing users to accept the update with a single button push. The code scanning intervention was implemented through automated scanning events and trigger code scanning by push or pull commands. GitHub carefully customized the interventions to suit the needs of software developers, taking advantage of existing tools, which had already been widely accepted by users.


\paragraph{\textbf{User recommendations}}
Each intervention provides recommendations on how to mitigate detected security issues. We have shown positive effects on the update behavior of vulnerable dependencies through security alerts and update interventions. Automated patching routines should be included in the code analysis intervention to improve its effectiveness. CodeQL's recommendations could offer source code examples that help or can be reused to fix identified issues. Potentially helpful code examples can be identified by searching for similar vulnerable code on GitHub that has been patched by other users. This approach has been shown to be very effective in the context of programming advice platforms~\cite{Fischer19}.


\paragraph{\textbf{Patch generation}}
The identified vulnerable source code and its updated counter examples in our dataset can be used to fine-tune large language models (LLMs) for automated patch generation. Fried et al. have recently developed an LLM that can perform infilling tasks on source code~\cite{fried2022incoder}. This model can learn from given example pairs how to fill masked parts of the source code. In this case, the vulnerable part of the source code would be masked, and the task of the model would be to fill it with the patched counterpart. Such a model could be used to suggest automated patches in a pull request, similar to the security update intervention, saving developers time and effort.


\section{Conclusion}
\label{sec:conclusions}

Our observations about GitHub's interventions align with several findings that have been seen in prior research. For instance, security alerts do have an effect on security (149.0\% gain in manual security updates), but are even more effective if they provide a seamless way out of the warning situation (additional 31.7\% gain in automated security updates). However, GitHub did not consistently apply these findings to all of their interventions since the code scanning intervention only provides alerts and is not active per default. Moreover, these alerts are difficult to recognize and detailed information about the found issues is not easily accessible. We assume that these inconsistencies in the design have led to a much smaller impact of the code scanning intervention (2\% gain in patched vulnerabilities). However, since the code scanning intervention is rather new and GitHub used a staged approach to improve their prior security interventions, we might see likewise improvements in the code scanning feature in the future.
 
\begin{acks}
We extend our thanks to Marisol Barrientos Moreno for her preliminary contributions to the GitHub analysis. We appreciate the support received from the Leibniz Supercomputing Centre of the Bavarian Academy of Sciences and Humanities. Additionally, we acknowledge the reviewers for their constructive comments, which have contributed to the improvement of this research paper.
\end{acks}

{\normalsize \bibliographystyle{plain}
\balance
\bibliography{refs}}

\begin{thebibliography}{10}

\bibitem{acar2017comparing}
Yasemin Acar, Michael Backes, Sascha Fahl, Simson Garfinkel, Doowon Kim, Michelle~L. Mazurek, and Christian Stransky.
\newblock Comparing the usability of cryptographic {API}s.
\newblock In {\em IEEE Symposium on Security and Privacy (S\&P)}, pages 154--171, 2017.

\bibitem{Acar:dev:2016}
Yasemin Acar, Michael Backes, Sascha Fahl, Doowon Kim, Michelle~L. Mazurek, and Christian Stransky.
\newblock You get where you're looking for: {T}he impact of information sources on code security.
\newblock In {\em IEEE Symposium on Security and Privacy (S\&P)}, pages 289--305, 2016.

\bibitem{Acar17Github}
Yasemin Acar, Christian Stransky, Dominik Wermke, Michelle~L. Mazurek, and Sascha Fahl.
\newblock Security developer studies with {G}ithub users: {E}xploring a convenience sample.
\newblock In {\em Thirteenth Symposium on Usable Privacy and Security (SOUPS)}, pages 81--95, 2017.

\bibitem{angermeir2021enterprise}
Florian Angermeir, Markus Voggenreiter, Fabiola Moyón, and Daniel Mendez.
\newblock Enterprise-driven open source software: A case study on security automation.
\newblock In {\em 2021 IEEE/ACM 43rd International Conference on Software Engineering: Software Engineering in Practice (ICSE-SEIP)}, pages 278--287, 2021.

\bibitem{bertrand2002much}
Marianne Bertrand, Esther Duflo, and Sendhil Mullainathan.
\newblock How much should we trust differences-in-differences estimation?
\newblock {\em The Quarterly Journal of Economics}, 119(1):249--275, 2004.

\bibitem{Borges18}
Hudson Borges and Marco~Tulio Valente.
\newblock What's in a {G}it{H}ub star? {U}nderstanding repository starring practices in a social coding platform.
\newblock {\em Journal of Systems and Software}, 146:112--129, 2018.

\bibitem{brodersen2015inferring}
Kay~H. Brodersen, Fabian Gallusser, Jim Koehler, Nicolas Remy, and Steven~L. Scott.
\newblock Inferring causal impact using {B}ayesian structural time-series models.
\newblock {\em The Annals of Applied Statistics}, 9(1):247--274, 2015.

\bibitem{chen2019reliable}
Mengsu Chen, Felix Fischer, Na~Meng, Xiaoyin Wang, and Jens Grossklags.
\newblock How reliable is the crowdsourced knowledge of security implementation?
\newblock In {\em 2019 IEEE/ACM 41st International Conference on Software Engineering (ICSE)}, pages 536--547, 2019.

\bibitem{danilova2020}
Anastasia Danilova, Alena Naiakshina, and Matthew Smith.
\newblock One size does not fit all: A grounded theory and online survey study of developer preferences for security warning types.
\newblock In {\em ACM/IEEE International Conference on Software Engineering (ICSE)}, pages 136--148, 2020.

\bibitem{fiebig2016one}
Tobias Fiebig, Anja Feldmann, and Matthias Petschick.
\newblock A one-year perspective on exposed in-memory key-value stores.
\newblock In {\em ACM Workshop on Automated Decision Making for Active Cyber Defense}, pages 17--22, 2016.

\bibitem{fischer2017stack}
Felix Fischer, Konstantin B{\"{o}}ttinger, Huang Xiao, Christian Stransky, Yasemin Acar, Michael Backes, and Sascha Fahl.
\newblock Stack {O}verflow considered harmful? {T}he impact of copy\&paste on {A}ndroid application security.
\newblock In {\em IEEE Symposium on Security and Privacy (S\&P)}, pages 121--136, 2017.

\bibitem{fischer2022nudging}
Felix Fischer and Jens Grossklags.
\newblock Nudging software developers toward secure code.
\newblock {\em IEEE Security \& Privacy}, 20(2):76--79, 2022.

\bibitem{fischer2021effect}
Felix Fischer, Yannick Stachelscheid, and Jens Grossklags.
\newblock The effect of {G}oogle search on software security: Unobtrusive security interventions via content re-ranking.
\newblock In {\em 2021 ACM SIGSAC Conference on Computer and Communications Security (CCS)}, pages 3070--3084, 2021.

\bibitem{Fischer19}
Felix Fischer, Huang Xiao, Ching-Yu Kao, Yannick Stachelscheid, Benjamin Johnson, Danial Razar, Paul Fawkesley, Nat Buckley, Konstantin B{\"o}ttinger, Paul Muntean, and Jens Grossklags.
\newblock Stack overflow considered helpful! {D}eep learning security nudges towards stronger cryptography.
\newblock In {\em 28th USENIX Security Symposium (USENIX Security)}, pages 339--356, 2019.

\bibitem{fried2022incoder}
Daniel Fried, Armen Aghajanyan, Jessy Lin, Sida Wang, Eric Wallace, Freda Shi, Ruiqi Zhong, Wen-tau Yih, Luke Zettlemoyer, and Mike Lewis.
\newblock In{C}oder: {A} generative model for code infilling and synthesis.
\newblock In {\em Eleventh International Conference on Learning Representations (ICLR)}, 2023.

\bibitem{Geiger17}
Stuart Geiger.
\newblock Summary analysis of the 2017 {G}it{H}ub open source survey.
\newblock {\em arXiv preprint arXiv:1706.02777}, 2017.

\bibitem{GitHub17}
{GitHub}.
\newblock Open source survey, 2017.
\newblock Available at: \url{https://opensourcesurvey.org/2017/}. Last accessed on: September 08, 2023.

\bibitem{gorski2018developers}
Peter~Leo Gorski, Luigi~Lo Iacono, Dominik Wermke, Christian Stransky, Sebastian M{\"o}ller, Yasemin Acar, and Sascha Fahl.
\newblock Developers deserve security warnings, too: On the effect of integrated security advice on cryptographic {API} misuse.
\newblock In {\em Fourteenth Symposium on Usable Privacy and Security (SOUPS)}, pages 265--281, 2018.

\bibitem{Gousios15}
Georgios Gousios, Andy Zaidman, Margaret-Anne Storey, and Arie Van~Deursen.
\newblock Work practices and challenges in pull-based development: {T}he integrator's perspective.
\newblock In {\em IEEE/ACM International Conference on Software Engineering (ICSE)}, pages 358--368, 2015.

\bibitem{Goyal18}
Raman Goyal, Gabriel Ferreira, Christian K{\"a}stner, and James Herbsleb.
\newblock Identifying unusual commits on {G}it{H}ub.
\newblock {\em Journal of Software: Evolution and Process}, 30(1):{Article No. e1893}, 2018.

\bibitem{Jiang17}
Jing Jiang, David Lo, Jiahuan He, Xin Xia, Pavneet~Singh Kochhar, and Li~Zhang.
\newblock Why and how developers fork what from whom in {G}it{H}ub.
\newblock {\em Empirical Software Engineering}, 22(1):547--578, 2017.

\bibitem{johnson13}
Brittany Johnson, Yoonki Song, Emerson Murphy-Hill, and Robert Bowdidge.
\newblock Why don't software developers use static analysis tools to find bugs?
\newblock In {\em ACM/IEEE International Conference on Software Engineering (ICSE)}, pages 672--681, 2013.

\bibitem{kavaler2019tool}
David Kavaler, Asher Trockman, Bogdan Vasilescu, and Vladimir Filkov.
\newblock Tool choice matters: {J}ava{S}cript quality assurance tools and usage outcomes in {G}it{H}ub projects.
\newblock In {\em 2019 IEEE/ACM 41st International Conference on Software Engineering (ICSE)}, pages 476--487, 2019.

\bibitem{kinsman2021how}
Timothy Kinsman, Mairieli Wessel, Marco~A. Gerosa, and Christoph Treude.
\newblock How do software developers use {G}it{H}ub actions to automate their workflows?
\newblock In {\em 2021 IEEE/ACM 18th International Conference on Mining Software Repositories (MSR)}, pages 420--431, 2021.

\bibitem{nguyen2017stitch}
Duc~Cuong Nguyen, Dominik Wermke, Yasemin Acar, Michael Backes, Charles Weir, and Sascha Fahl.
\newblock A stitch in time: {S}upporting {A}ndroid developers in writing secure code.
\newblock In {\em ACM Conference on Computer and Communications Security (CCS)}, pages 1065--1077, 2017.

\bibitem{panichella2021wont}
Sebastiano Panichella, Gerardo Canfora, and Andrea {Di Sorbo}.
\newblock ``{W}on’t we fix this issue?'' {Q}ualitative characterization and automated identification of wontfix issues on {G}it{H}ub.
\newblock {\em Information and Software Technology}, 139:{Article No. 106665}, 2021.

\bibitem{pearce2022asleep}
Hammond Pearce, Baleegh Ahmad, Benjamin Tan, Brendan Dolan-Gavitt, and Ramesh Karri.
\newblock Asleep at the keyboard? {A}ssessing the security of {G}it{H}ub {C}opilot's code contributions.
\newblock In {\em 2022 IEEE Symposium on Security and Privacy (S\&P)}, pages 980--994, 2022.

\bibitem{Pinto16}
Gustavo Pinto, Igor Steinmacher, and Marco~Aur{\'e}lio Gerosa.
\newblock More common than you think: {A}n in-depth study of casual contributors.
\newblock In {\em 2016 IEEE 23rd International Conference on Software Analysis, Evolution, and Reengineering (SANER)}, pages 112--123, 2016.

\bibitem{Ramos16}
Miguel Ramos, Marco~Tulio Valente, Ricardo Terra, and Gustavo Santos.
\newblock Angular{JS} in the wild: {A} survey with 460 developers.
\newblock In {\em 7th International Workshop on Evaluation and Usability of Programming Languages and Tools}, pages 9--16, 2016.

\bibitem{Rastogi16}
Ayushi Rastogi.
\newblock Do biases related to geographical location influence work-related decisions in {G}it{H}ub?
\newblock In {\em IEEE/ACM International Conference on Software Engineering Companion (ICSE-C)}, pages 665--667, 2016.

\bibitem{reeder2018experience}
Robert~W Reeder, Adrienne~Porter Felt, Sunny Consolvo, Nathan Malkin, Christopher Thompson, and Serge Egelman.
\newblock An experience sampling study of user reactions to browser warnings in the field.
\newblock In {\em Proceedings of the 2018 CHI conference on human factors in computing systems}, pages 1--13, 2018.

\bibitem{Saito16}
Yusuke Saito, Kenji Fujiwara, Hiroshi Igaki, Norihiro Yoshida, and Hajimu Iida.
\newblock How do {G}it{H}ub users feel with pull-based development?
\newblock In {\em 2016 7th International Workshop on Empirical Software Engineering in Practice (IWESEP)}, pages 7--11. IEEE, 2016.

\bibitem{takhteyev2010investigating}
Yuri Takhteyev and Andrew Hilts.
\newblock Investigating the geography of open source software through {G}it{H}ub.
\newblock Technical report, University of Toronto, 2010.

\bibitem{tan2021github}
Shin~Hwei Tan, Chunfeng Hu, Ziqiang Li, Xiaowen Zhang, and Ying Zhou.
\newblock Git{H}ub-{OSS} {F}ixit: {F}ixing bugs at scale in a software engineering course.
\newblock In {\em 2021 IEEE/ACM 43rd International Conference on Software Engineering: Software Engineering Education and Training (ICSE-SEET)}, pages 1--10, 2021.

\bibitem{Vasilescu15}
Bogdan Vasilescu, Vladimir Filkov, and Alexander Serebrenik.
\newblock Perceptions of diversity on {G}it{H}ub: {A} user survey.
\newblock In {\em 2015 IEEE/ACM 8th International Workshop on Cooperative and Human Aspects of Software Engineering}, pages 50--56, 2015.

\bibitem{Vasilescu15b}
Bogdan Vasilescu, Daryl Posnett, Baishakhi Ray, Mark van~den Brand, Alexander Serebrenik, Premkumar Devanbu, and Vladimir Filkov.
\newblock Gender and tenure diversity in {G}it{H}ub teams.
\newblock In {\em 33rd Annual ACM Conference on Human Factors in Computing Systems (CHI)}, pages 3789--3798, 2015.

\bibitem{Wu14}
Yu~Wu, Jessica Kropczynski, Patrick Shih, and John Carroll.
\newblock Exploring the ecosystem of software developers on {G}it{H}ub and other platforms.
\newblock In {\em Companion Publication of the 17th ACM Conference on Computer Supported Cooperative Work \& Social Computing (CSCW Companion)}, pages 265--268, 2014.

\end{thebibliography}

\ifthenelse{\boolean{isArxiv}}{
\appendix
\section{Appendix: Supporting Figures}
We present the UI and user journey for each intervention in Figures~\ref{fig:security_ui}, \ref{fig:udpate_ui} and~\ref{fig:codeql_ui}.
\begin{figure*}[t]
\centering
\includegraphics[width=1.0\textwidth]{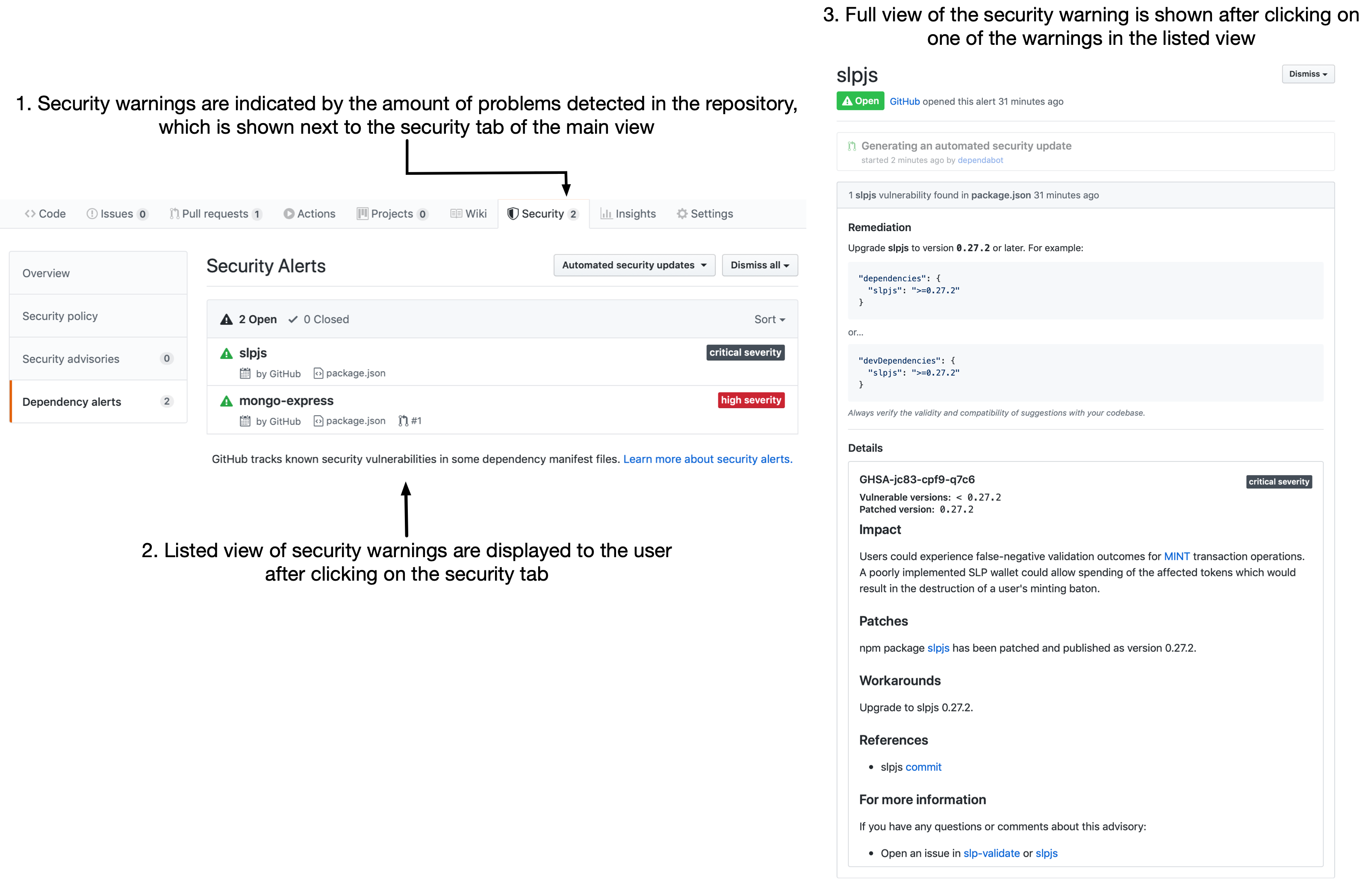} 
\caption{UI and user journey of the security alert intervention}
\label{fig:security_ui}
\end{figure*}

\begin{figure*}[t]
\centering
\includegraphics[width=1.0\textwidth]{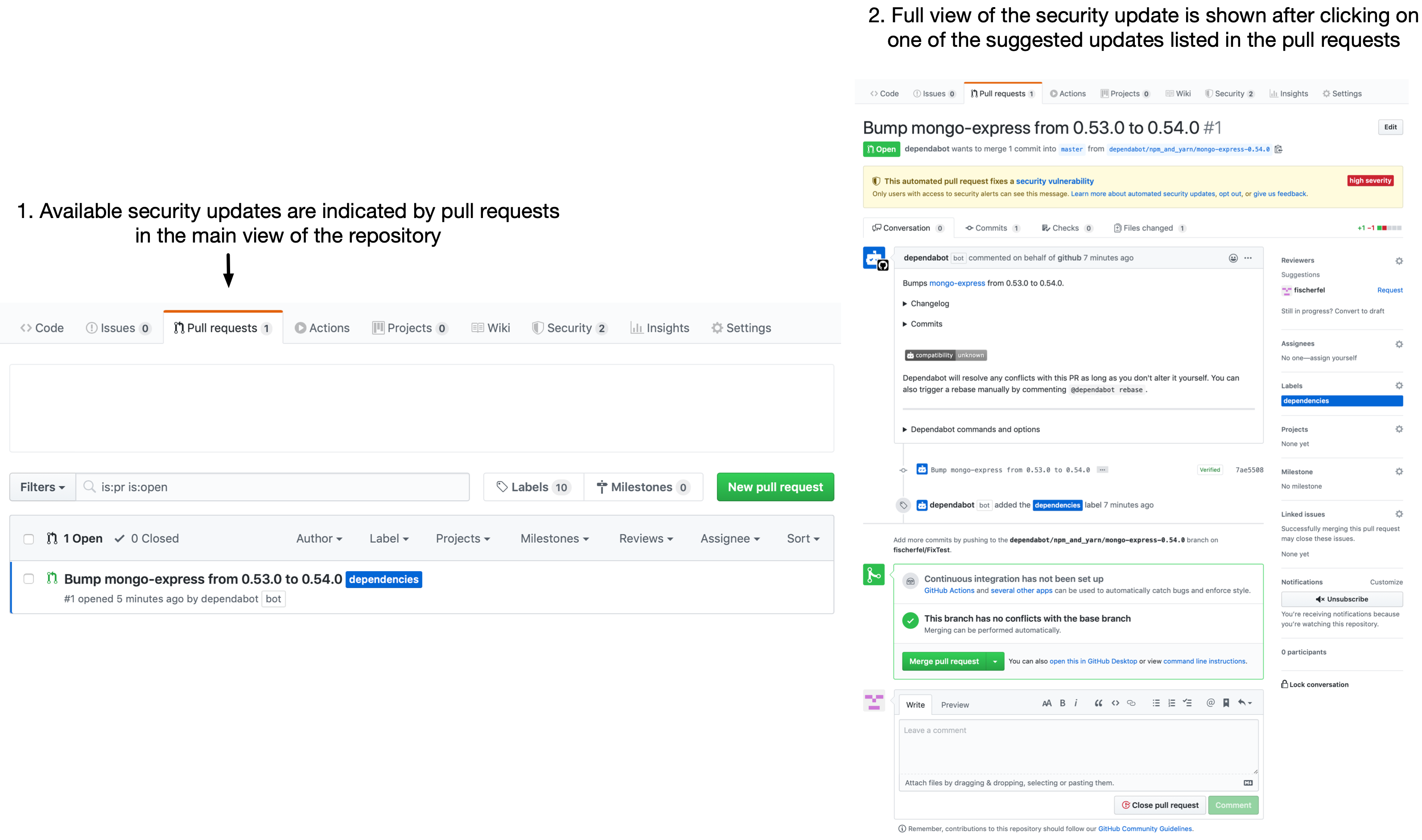} 
\caption{UI and user journey of the security update intervention}
\label{fig:udpate_ui}
\end{figure*}

\begin{figure*}[t]
\centering
    \begin{subfigure}[t]{0.5\textwidth}
        \includegraphics[width=1.0\textwidth]{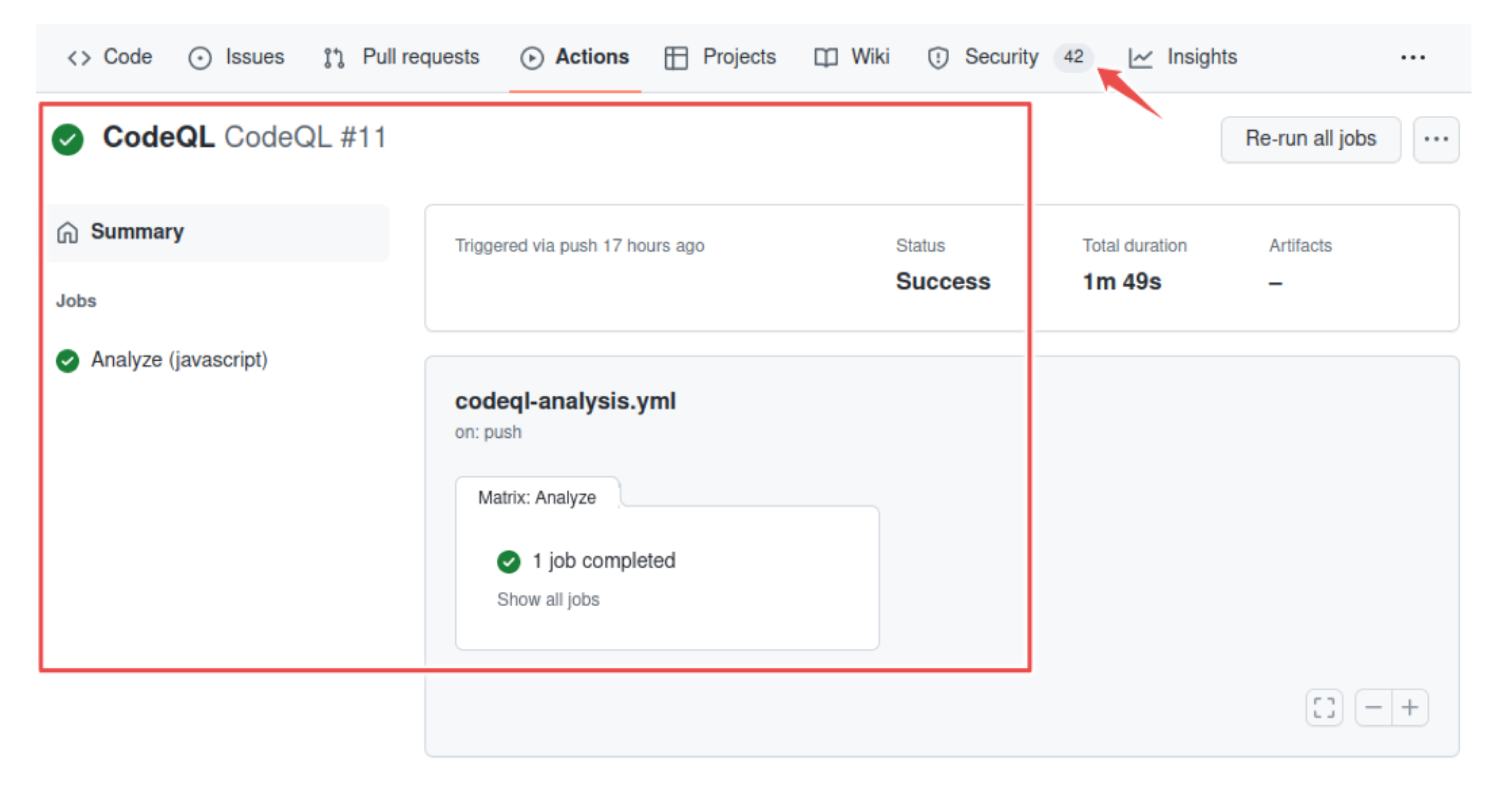} 
    \end{subfigure}%
    \begin{subfigure}[t]{0.5\textwidth}
        \includegraphics[width=1.0\textwidth]{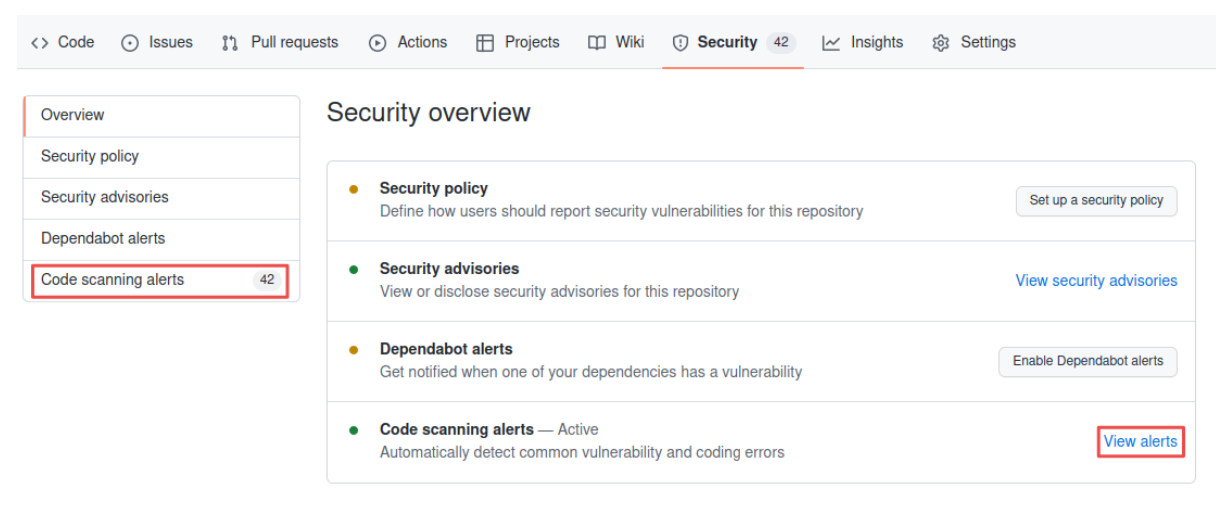}
    \end{subfigure}
\caption{The left screenshot shows GitHub's security indicator if CodeQL found a vulnerability in the source code of the project. The right screenshot shows the web page path the user has to follow to view the code scanning results.}
\label{fig:codeql_ui}
\end{figure*}
}{}

\end{document}